\documentclass[sigplan, nonacm]{acmart}

\usepackage{algorithm}  % 提供浮动体环境
\usepackage{algpseudocode}  % 提供伪代码命令
\renewcommand\footnotetextcopyrightpermission[1]{}
\usepackage{xspace} % for \xspace command
\usepackage{booktabs}
\usepackage{multirow}
\usepackage{subcaption} % for \subcaption command

\newcommand{\sys}{\texttt{shadowAttn}\xspace}

\usepackage{tikz}           % TikZ 画图支持

\newcommand{\circled}[1]{%
  \tikz[baseline=(char.base)]{%
    \node[shape=circle, draw, fill=black, inner sep=0.7pt, text=white] (char) {#1};%
  }%
}

\begin{document}
\title{ShadowNPU: System and Algorithm Co-design for NPU-Centric On-Device LLM Inference}

\begin{abstract}

% \mwx{
% On-device LLM has become a critical feature towards ubiquitous mobile intelligence, enabling novel applications such as personal assistants.
% Ideally, LLM inference shall be dedicated to mobile NPUs for extreme energy efficiency and fast processing; in practice, we find the attention operation often falls back to mobile CPU/GPU due to high accuracy loss when running on NPUs' integer processing unit.
% Such falling back not only compromises the overall energy efficiency, but also leads to severe resource contention with other workloads co-located on CPU/GPU.
% Thereby, a key design challenge of on-device LLM system is to guarantee the generation accuracy and speed, while consuming as few as possible the FP-based processing.
% }

On-device running Large Language Models (LLMs) is nowadays a critical enabler towards preserving user privacy.
We observe that the attention operator falls back from the special-purpose NPU to the general-purpose CPU/GPU because of quantization sensitivity in state-of-the-art frameworks.
This fallback results in a degraded user experience and increased complexity in system scheduling.
To this end, this paper presents \sys, a system-algorithm codesigned sparse attention module with minimal reliance on CPU/GPU by only sparsely calculating the attention on a tiny portion of tokens.
The key idea is to hide the overhead of estimating the important tokens with a NPU-based pilot compute.
Further, \sys proposes insightful techniques such as NPU compute graph bucketing, head-wise NPU-CPU/GPU pipeline and per-head fine-grained sparsity ratio to achieve high accuracy and efficiency.
\sys delivers the best performance with highly limited CPU/GPU resource; it requires much less CPU/GPU resource to deliver on-par performance of SoTA frameworks.

\end{abstract}

\author{Wangsong Yin$^{\blacklozenge}$, Daliang Xu$^{\Diamond}$, Mengwei Xu$^{\Diamond}$,
Gang Huang$^{\blacklozenge}$, Xuanzhe Liu$^{\blacklozenge\#}$}
\affiliation {
	\institution{$^\blacklozenge$Key Lab of High Confidence Software Technologies (Peking University), Beijing, China}
	\country{}
}

\affiliation {
	\institution{$^\Diamond$State Key Laboratory of Networking and Switching Technology (BUPT), Beijing, China}
	\country{}
}

\email{yws@stu.pku.edu.cn}
% \email{
% 	mwx@bupt.edu.cn
% }
% \email{
% liuxuanzhe@pku.edu.cn
% }

\thanks{$^\#$Corresponding author.}

\maketitle % should come after the abstract

% add the paper content here

\section{Introduction}

% Tell a story of accelerator-only on-device inference.

% 1. 未来要用NPU做大模型，CPU和GPU不适合
% 2. 现有框架attention会fall back to cpu/gpu的FP计算，因为COTS NPU只有非常有限的sparse FP算力支持，导致attention的精度损失过大。
% 3. fall back会有两层问题，（1） efficiency （2） 推理不稳定
% 4. 本文的目标是propose一个system-algo codesign that achieves best performance under weak FP compute ability。在COTS NPU上，这个系统的计算

% (1) Existing frameworks' attention will fall back to CPU/GPUs that are outside the accelerator.

% (2) This is not good, as CPU/GPU is of other use.

% (3) The reason why attention cannot run on NPUs...

% (4) Our vision: introducing a new co-processor, Float Sparse Unit (FSU) into the next-generation NPUs.

% (5) Many challenges to tackle. To run efficient and accurate sparse attention on such NPUs, we need to solve xxx.

% (6) \sys: a system-algorithm codesign on top of our visioned next-generation mobile NPUs.

% \yws{IDU: Int Dense Units; FSU: Float Sparse Units}

% (1) para1: on-device llm and attention module

% (2) para2: mobile SoC and inference frameworks

% (3) para3: the fall back of attention, why

% (4) para4: the problem of fall back, NPU-centric

% (5) para5: our system, challenges

% (6) para6: design

On-device Large Language Models (LLMs) has been a critical enabler to privacy-preserving and ubiquitous artificial intelligence.
Recently, the emergence of mobile-size LLMs like Qwen series~\cite{bai2023qwentechnicalreport, qwen2025qwen25technicalreport} catalyzes killer applications such as content summarization and GUI/API agents~\cite{rewind, chen2024octopusv2ondevicelanguage, xie2024droidcalldatasetllmpoweredandroid, zhang2024llamatouch}.

\noindent \textbf{NPU-centric LLM inference.}
Mobile devices typically incorporate heterogeneous processors such as CPU, GPU, and NPU.
Ideally, the LLM inference shall be end-to-end executed on the mobile NPUs with minimal out-of-NPU resources.
The advantages of such NPU-centric LLM inference are twofold.
First, mobile NPUs, as dedicated for neural networks, are with high-throughput integer capability and much more power-saving than CPU/GPU~\cite{llm.npu, 10.1145/3689031.3696067, chen2025heterollmacceleratinglargelanguage, NN-Stretch}.
Second, it avoids resource contention with other general-purpose mobile workloads (e.g., user interaction handling and rendering) that rely on CPU and GPU.
Co-locating LLM inference with those workloads can easily hamper user experience, adding extra complexity to the system scheduling~\cite{10.1145/3372224.3419192, 10.1145/3498361.3538948, 9796661, zou2025surveyrealtimeschedulingacceleratorbased, 10.1145/3676641.3716278}.

However, in practice we find the attention operation in LLMs often falls back to mobile CPU/GPU in state-of-the-art on-device inference frameworks~\cite{llm.npu, mlc-llm, chen2025heterollmacceleratinglargelanguage, llamacpp}.
For instance, llm.npu~\cite{llm.npu} offloads the non-attention operations to NPUs; yet the attention still runs on CPUs --- not NPU-centric.

% \xdl{Mismatch between NPUs' design and attention.}
Such a compromise is underpinned by the rationale of quantization sensitivity.
% \xdl{1. quantization; 2. dynamic shape (parameters). Merge the the limitations of quantization to the first point.}
% \textit{(1) The quantization sensitivity of attention operation.}
On one hand, activations in LLMs feature much more hard-to-quantize outliers compared to weights~\cite{xiao2024smoothquantaccurateefficientposttraining, lin2024awqactivationawareweightquantization}, 
while the attention operation computes on multiple activation tensors, i.e., the Q, K, V tensors.
On the other hand, the NPU's static graph hampers fine-grained quantization.
It sets superparameters (tensor shape and constants such as scale factor) before compilation for resource scheduling.
Such a paradigm leads to sharing a fixed scale factor for an entire tensor, further limiting the attention quantization. 
% \textit{(2) The limitations of quantization on NPUs.}
% At the hardware level, fine-grained quantization methods such as per-group quantization fail to fully utilize the throughput of NPUs, leading to significant performance degradation.
% Thus, the NPU typically employs a per-tensor quantization, which shares one scale factor across an entire tensor.
% At the software level, the static graph sets superparameters (tensor shape and constants such as scale factor) before compilation for resource scheduling.
% Such a paradigm requires the scale factor to be fixed, further limiting the attention quantization.
In the following experiments (Table~\ref{tab:cpuvsnpu}), we find that NPU-based attention leads to on average 18 pp accuracy drop on mobile LLMs/tasks.
This echos the necessity of offloading the attention to CPU/GPU.

\noindent \textbf{NPU-centric LLM inference by sparse attention.}
% However, there exists a crucial problem in current CPU/GPU and NPU collaborated paradigm: it leads to severe resource contention between other apps' workloads co-located on CPU/GPU.
% For instance, when watching 60 FPS 1080P videos together with running the navigation service, only 1--2 CPU cores out of the CPU and GPU resources are left for LLM inference on MI14 smartphone.
% \xdl{energy efficiency problem.}
% Such a limitation calls for \textit{NPU-centric inference}, which means completely offloading the inference to NPUs and minimizing the resource consumption of CPU/GPU.
A key opportunity to minimize the reliance on CPU/GPU resource is the the highly sparse feature of attention operations.
A small portion of tokens can be much more important than others; sparsely computing these tokens can significantly reduce the CPU/GPU computation with almost no accuracy drop.
As detailed in $\S$\ref{sec:bkg}, in the Qwen2-1.5B model, on average more than 80\% of the tokens are assigned relatively low importance in the WikiText-2 dataset.

Surprisingly, despite the promised advantages, our preliminary results indicate that directly applying sparse attention in LLM inference yields no performance gain, as before the sparse computation can be executed, an estimation stage is required to evaluate token importance, incurring substantial overhead on the CPU/GPU. This stage involves a matrix multiplication to compute attention scores for each token in $Q$ and $K$~\cite{zhang2023h2oheavyhitteroracleefficient, wang2021spatten}. Consequently, the estimation stage dominates the overall attention computation under high sparsity. For instance, with a sparsity ratio of $20\%$ (i.e., $80\%$ of tokens pruned), estimation alone contributes over $60\%$ of the total overhead, reducing CPU/GPU resource consumption by only about $20\%$.
Intuitively, we can downsample the Q and K for an efficient estimation, yet leading to significant accuracy drop.
For instance, when eliminating the estimation overhead by grouping adjacent tokens in a block and only computing the block-level importance, the accuracy undergoes an average drop of 4 pp on mobile LLMs/tasks compared to token level sparsity.
To this end, the key question of \sys's design is how to compute sparse attention accurately and efficiently in NPU-centric LLM inference.

\noindent \textbf{\sys: Dynamic sparse attention with NPU-based estimation.}
In this paper, we propose \sys, a system-algorithm co-designed on-device sparse attention module for NPU-centric LLM inference while minimizing its reliance on CPU/GPU fallback (e.g., only one CPU core).
Its key idea is that the estimation of important tokens can be much more resilient to quantization compared to the end-to-end attention.
The rationale is that determining the important tokens only requires the relative value of attention scores, while calculating the attention's exact result requires the absolute value.
Thus, \sys offloads the estimation to NPU, and transfers the position indices of the important tokens to CPU/GPU for further sparse attention calculation.
By doing so, only a small portion of tokens are computed on CPU/GPU with high precision float operations.
Besides, \sys also identifies the uneven property of each head's sparsity ratio.
It determines the ratio for each head by a lightweight offline profiling, further optimizing the accuracy-efficiency tradeoff.
When designing \sys, however, we need further
to address the following major challenges that have not been explored in the existing literature.

% \noindent \textbf{Challenges.}
% It minimizes the reliance on the CPU/GPU compute by identifying and tackling the following challenges.

% \textit{(1) The heavy overhead of accurate token importance estimation.}
% To estimate the importance of tokens, each token in Q and K needs to be computed by a matrix multiplication operation to get the attention scores~\cite{zhang2023h2oheavyhitteroracleefficient, wang2021spatten}.
% The challenge is that an accurate estimation leads to heavy overhead on CPU/GPU.
% On one hand, calculating each token in estimation can even consume more compute than the sparse attention after acquiring the important tokens.
% For instance, the estimation incurs over 60\% overhead under a sparsity ratio of 20\% (i.e., 80\% tokens are sparsed).
% Such a highly sparse attention only reduces the resource reliance on CPU/GPU by about 20\%.
% On the other hand, simply downsampling the Q and K for an efficient estimation leads to significant accuracy drop.
% For instance, when eliminating the estimation overhead by grouping adjacent tokens in a block and only computing the block-level importance, the accuracy undergoes an average drop of 4 pp on mobile LLMs/tasks compared to token level sparsity.
% To this end, \sys's key idea is that the estimation can be offloaded to NPU to fully utilize its integer dense compute.
% We elaborate the rationale in the the following design details. 
% % \xdl{It is better to move these two sentences directly to design.}

\textit{(1) The inflexibility of NPU static compute graph.}
The software stack of NPUs employs a static compute graph, which specifies the shape and quantization scale factors as constant for offline optimization.
However, the attention input tensors are highly dynamic;
their scale factors exhibit large fluctuation.
The tensor shape may also vary in different cases.
Directly generating one static compute graph for the attention leads to accuracy drop and under utilization of NPU.

\textit{(2) Multi-faceted NP-hard NPU–CPU/GPU scheduling. }
% \xdl{Multi-faceted NP-hard NPU–CPU/GPU scheduling.}
% \sys's sparse attention involves both the NPU and CPU/GPU.
% % Naively running the heterogeneous workloads sequentially overlooks the opportunity of pipeline execution.
% They can be overlapped by a head-level fine-grained pipeline execution.
% On one hand, the pipeline can be optimized from various deeply buried aspects like NPU kernel fused launching.
% Its performance may be degraded without carefully going through all of these aspects.
% On the other hand, the pipeline planning is NP-hard.
% Solving such problem on mobile devices with minimal overhead is also challenging.
The two stages of \sys's sparse attention can be overlapped by a head-level fine-grained pipeline on both the NPU and CPU/GPU. However, achieving efficient fine-grained scheduling is non-trival and challenging because (i) the pipeline can be optimized from various deeply buried aspects like NPU kernel fused launching.
Its performance may be degraded without carefully going through all of these aspects. (ii)  the pipeline planning problem is NP-hard.

% Characterizing the overlap-able parts of workload and planning a high performance execution pipeline are non-trivial challenges.
% \xdl{emphasize naive pipeline challenges or mention NPU and CPU/GPU cannot pipeline without head-level analysis.}

% \noindent \textbf{Techniques.}
\sys incorporates the following techniques with deep insights towards the aforementioned challenges.

% \textit{(1) Dynamic sparse attention with NPU-based estimation} ($\S$\ref{subsec:sparse_attn}).
% \sys's sparse attention is tailored for both the accuracy and efficiency.
% Its key observation is that the estimation of important tokens can be much more resilient to quantization compared to the end-to-end attention.
% The rationale is that determining the important tokens only requires the relative value of attention scores, while calculating the attention's exact result requires the absolute value.
% Thus, \sys offloads the estimation to NPU, and transfers the position indices of the important tokens to CPU/GPU for further sparse attention calculation.
% By doing so, only a small portion of tokens are computed on CPU/GPU with high precision float operations.
% Besides, \sys also identifies the uneven property of each head's sparsity ratio.
% It determines the ratio for each head by a lightweight offline profiling, further optimizing the accuracy-efficiency tradeoff.  

\textit{(1) NPU compute graph bucketing} ($\S$\ref{subsec:buckets}).
To mitigate the limitations of static compute graph, \sys offline generates multiple graphs and organizes them in buckets for online selection.
Specifically, a bucket contains graphs with various shapes and the same scale factor of input tensors.
At online inference stage, the bucket whose scaling factor is closest to that of the input tensor is selected, followed by selecting a graph with an appropriate shape from the bucket.

\textit{(2) Head-wise NPU-CPU/GPU pipeline} ($\S$\ref{subsec:pipeline}).
\sys proposes a head-level pipeline that leverages the following insights.
Firstly, it overlaps the NPU estimation, CPU/GPU top k operation and CPU/GPU sparse attention with each other.
On top of this, it launches the NPU kernels with the same scale in one-shot to maximize the NPU utilization, and carefully plans the execution order of each head to minimize the pipeline bubbles.
To achieve fast on-device planning, it simplifies the NP-hard planning complexity by a heuristic greedy search, which selects the head that minimizes the pipeline latency at each step.

\noindent \textbf{Implementation and Evaluation.}
We prototype \sys in over 10,000 LoC of C++/Python. 
We test \sys on commercial-off-the-shelf smartphones MI14 and Redmi K60 Champion Edition with snapdragon 8gen3/8gen2 equipped.
% \sys is compatible with any on-device LLM
% inference library on mainstream mobile/edge SoCs. 
% It can be integrated by directly including a header file, being least intrusive to the original codebase.
We only use one CPU core for necessary control flow and sparse compute, with other compute completely on NPUs.

We test both the breakdown performance of the attention module and the end-to-end performance of integrating \sys into the state-of-the-art NPU framework llm.npu~\cite{llm.npu}.
The evaluation is conducted on mobile LLMs, including Qwen2-0.5B/1.5B and PhoneLM-0.5B/1.5B.
We report the results on three representative datasets of mobile-specific tasks (NLP and agents).
Compared to four design alternatives with a circumstance of highly limited CPU/GPU resources, \sys exhibits up to 6.9$\times$ breakdown speedup, up to 4.5$\times$ end-to-end speedup and up to 7.7$\times$ energy reduction.
Compared to native attention of SoTA frameworks, \sys achieves on-par or even better performance with significantly fewer CPU/GPU resources.
The accuracy loss is only 0.4 pp on average on four models and three datasets.
% To the best of our knowledge, for the first time \sys enables the on-device LLM inference to run together with common mobile apps such as watching videos and browsing shopping apps, with no obvious performance degradation.

\noindent \textbf{Contributions} are as follows.

$\bullet$ We identify the significant problem of NPU-centric LLM inference on mobile devices, and propose \sys, a system-algorithm codesigned sparse attention that significantly reduces the reliance on float compute of CPU/GPU.

$\bullet$ We propose several key techniques, including NPU offloading of token importance estimation, NPU compute graph bucketing and NPU-CPU/GPU pipeline, and tackle the main challenges of NPU-centric on-device attention.

$\bullet$ \sys outperforms strong baselines while minimizing CPU/GPU resource. \sys can be plug-and-play to any mainstream on-device LLM inference frameworks.
\section{Background and Motivation}
\label{sec:bkg}

\subsection{LLM Inference on Mobile SoCs}

\begin{table}[]
\footnotesize
\begin{tabular}{c|c|c|c|c}
\hline
\textbf{Vendor} &
  \textbf{SoC} &
  \textbf{NPU} &
  \textbf{\begin{tabular}[c]{@{}c@{}}INT \\ Capability\end{tabular}} &
  \textbf{\begin{tabular}[c]{@{}c@{}}Float \\ Capability\end{tabular}} \\ \hline
Qualcomm & Snapdragon & \begin{tabular}[c]{@{}c@{}}Hexagon \\ DSP\end{tabular} & High & No  \\ \hline
Qualcomm & Snapdragon & HTP                                                    & High & Low \\ \hline
\begin{tabular}[c]{@{}c@{}}Texas \\ Instruments\end{tabular} &
  TMS320F2812 &
  \begin{tabular}[c]{@{}c@{}} C28x\\ DSP\end{tabular} &
  High &
  No \\ \hline
Nvidia   & Orin       & DLA                                                    & High & No  \\ \hline
Tesla    & FSD        & FSD D1                                                 & High & No  \\ \hline
\end{tabular}
\caption{NPUs of mainstream mobile/embedded SoCs. High-throughput integer capability is the main focus of NPUs.}
\end{table}

\begin{table}[]
\footnotesize
\begin{tabular}{c|c|c|c|c}
\hline
\textbf{Framework}  & llm.npu & HeteroLLM & llama.cpp & mlc-llm \\ \hline
\textbf{Atten.}     & CPU     & GPU       & CPU       & GPU     \\ \hline
\textbf{Non-Atten.} & NPU     & GPU/NPU   & CPU       & GPU     \\ \hline
\end{tabular}
\caption{On-device frameworks typically run the attention operation on CPU/GPUs.}
\label{tab:npus_in_socs}
\end{table}

\noindent \textbf{Mobile NPUs: integer accelerators for DNN inference.}
Due to increasing concerns about individual data protection and safety, more users are opting to run large language models directly on their personal devices rather than sending sensitive information to remote servers.
% To this end, NPUs are typically integrated into the mobile SoCs and are dedicated for DNN inference.
To achieve higher performance, on-device LLM inference shall be executed on NPUs, specialized hardware designed for DNN inference. Such paradigm is referred as \textit{NPU-centric inference}.
Table~\ref{tab:npus_in_socs} summarizes the NPUs on mainstream mobile/embedded SoCs, including Qualcomm Snapdragon series~\cite{8gen3, snpe}, Texas Instruments TMS320F2812~\cite{TMS320F2812}, Nvidia Orin~\cite{orin}, etc.
We observe that the commonality among these NPUs is their primary focus on \textit{low-precision integer compute}, i.e., all of them support integer operation, but have zero or relatively low float capabilities.
This situation arises because (i) DNNs are able to be quantized to lower precisions for efficient execution and memory saving compared to general workloads.
(ii) supporting multiple data precision types consumes more chip area or introduces significant additional design complexity for unit reusing~\cite{9916240, lokhande2025polaronprecisionawareondevicelearning, lokhande2024flexpeflexiblesimdmultiprecision}.

\begin{figure}[t]
    \centering
    \includegraphics[width=0.47\textwidth]{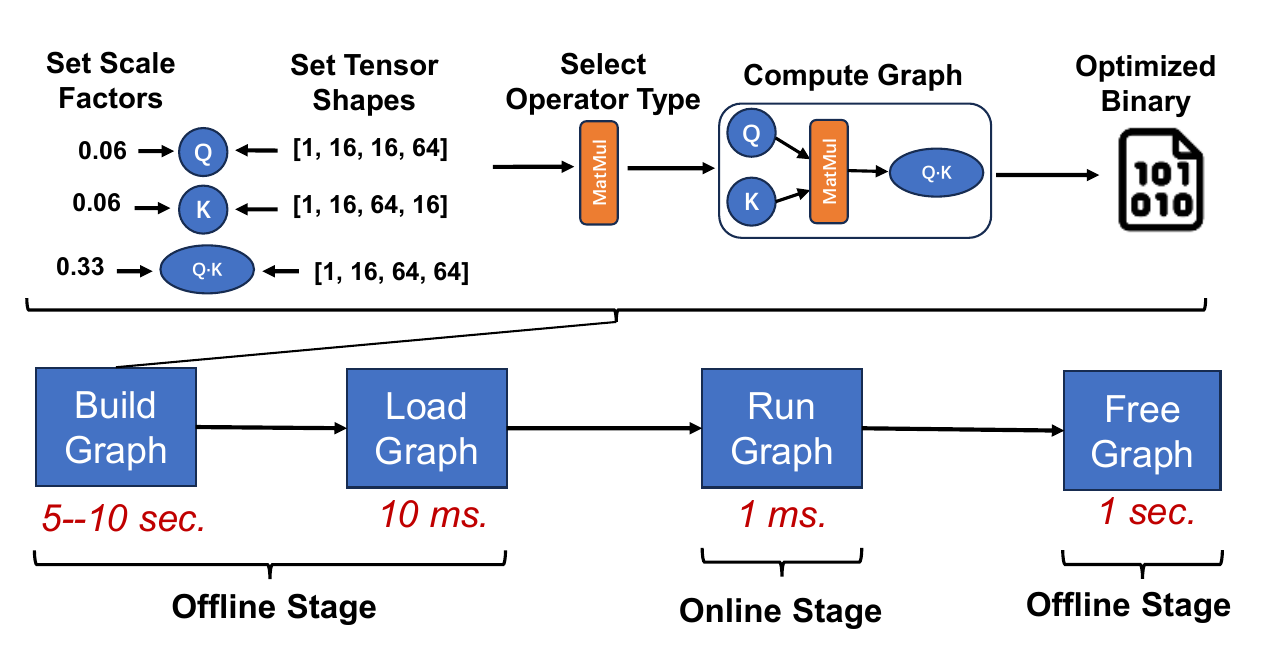}
    \caption{The workflow of static compute graph of mobile NPUs. The latency is acquired on QNN SDK~\cite{qnn} by a basic matrix multiplication operation.}
    \label{fig:static_graph}
\end{figure}

\noindent \textbf{The static compute graph of mobile NPUs.}
To schedule the resources for the best runtime efficiency, mobile NPUs typically employ a static compute graph.
Figure~\ref{fig:static_graph} illustrates the workflow of a static-graph-based inference procedure for QNN SDK~\cite{qnn}.
In the offline stage, a computation graph is first built and then loaded into the device memory.
During the online inference stage, the input data is fed into the graph for execution.
Finally, the graph is offline released from the memory.

The static graph requires several key hyperparameters to be specified and fixed during the offline graph-building stage.
As illustrated in Figure~\ref{fig:static_graph}, this procedure includes assigning a single float value as the scale factor for an INT8 tensor and defining the tensor exact shape.
The tensors and an operator type provided by the vendor are then assembled into a compute graph.
Finally, the graph is compiled into a binary file through a series of optimization passes.

The time overhead of each stage is listed in Figure~\ref{fig:static_graph}.
The offline stage is extremely time-consuming compared to the online stage.
This makes it impossible to dynamically online build the graph for each input.

% \usepackage{multirow}
\begin{table}[]
\footnotesize
\begin{tabular}{ccccccccc}
\hline
\multirow{2}{*}{\textbf{Dataset}} &
  \multicolumn{2}{c}{\textbf{\begin{tabular}[c]{@{}c@{}}PhoneLM\\ -0.5B\end{tabular}}} &
  \multicolumn{2}{c}{\textbf{\begin{tabular}[c]{@{}c@{}}PhoneLM\\ -1.5B\end{tabular}}} &
  \multicolumn{2}{c}{\textbf{\begin{tabular}[c]{@{}c@{}}Qwen2\\ -0.5B\end{tabular}}} &
  \multicolumn{2}{c}{\textbf{\begin{tabular}[c]{@{}c@{}}Qwen2\\ -1.5B\end{tabular}}} \\ \cline{2-9} 
          & \textbf{C/G} & \textbf{N} & \textbf{C/G} & \textbf{N} & \textbf{C/G} & \textbf{N} & \textbf{C/G} & \textbf{N} \\ \hline
ArxivSum~\cite{cohan-etal-2018-discourse}  & 14.7         & 0.0        & 11.9         & 0.0        & 10.7         & 9.4        & 8.5          & 9.1        \\
DroidCall~\cite{xie2024droidcalldatasetllmpoweredandroid} & 27.5         & 20.5       & 20.5         & 19.0       & 34.5         & 27.5       & 48.0         & 22.5       \\
Octopus~\cite{chen2024octopusv2ondevicelanguage}   & 64.6         & 24.1       & 79.2         & 24.7       & 60.6         & 34.8       & 61.2         & 34.2       \\ \hline
\end{tabular}
\caption{Accuracy on mobile LLMs and tasks. ``C/G'' means running attention on CPU/GPU in float32; ``N'' means running attention on NPU in INT8.}
\label{tab:cpuvsnpu}
\end{table}

\noindent \textbf{The attention falls back to CPU/GPUs in state-of-the-art on-device LLM inference frameworks.}
Table~\ref{tab:npus_in_socs} lists representative on-device LLM inference frameworks, including llm.npu~\cite{llm.npu}, HeteroLLM~\cite{chen2025heterollmacceleratinglargelanguage}, llama.cpp~\cite{llamacpp} and mlc-llm~\cite{mlc-llm}.
\textit{They all delegate the attention operation to CPU or GPU.}
For instance, although llm.npu offloads its non-attention parts to NPU, the attention operation still runs on CPU; HeteroLLM also employs such an NPU-CPU/GPU collaborated paradigm.
The CPU-based framework llama.cpp and GPU-based framework mlc-llm also show this property.
This is mainly because that the attention is unfriendly to mobile NPUs.
The attention compute involves multiple activations (i.e., the non-weight tensors in a neuron network), which are hard to quantize~\cite{xiao2024smoothquantaccurateefficientposttraining, lin2024awqactivationawareweightquantization}.
Further, the static graph limits the quantization on NPUs to per-tensor static quantization.
Such a quantization method is very coarse-grained. 
It shares a fixed scale factor across an entire tensor, preventing the NPU from adopting fine-grained methods such as K-quant~\cite{llamacpp} or AWQ~\cite{lin2024awqactivationawareweightquantization}.
Table~\ref{tab:cpuvsnpu} shows that offloading the attention operation to mobile NPUs leads to significant accuracy degradation. 
For instance, PhoneLM-0.5B shows a 14.7 pp accuracy drop and a 7 pp accuracy drop on ArxivSum dataset and DriodCall dataset, respectively.
On three datasets and four models, the NPU-based attention exhibits an average drop of 18 pp.

The reliance on CPU/GPU fallback for high accuracy makes the on-device inference not NPU-centric.
It fails to leverage the high-throughput compute capability and low energy consumption of mobile NPUs.
Even worse, it leads to severe resource contention between other mobile apps co-located on CPU/GPU.

\subsection{Sparse Attention}
The opportunity of minimizing the reliance on CPU/GPU is the highly sparse characteristic of attention operation.

\begin{figure}[t]
    \centering
    \includegraphics[width=0.48\textwidth]{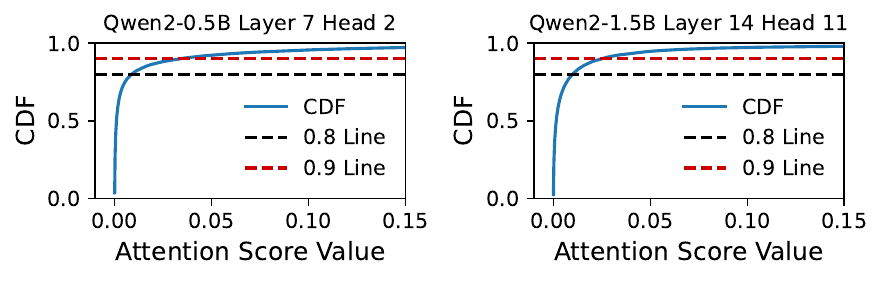}
    \caption{The attention score skewness of LLMs. Profiled on 128 samples from WikiText-2.}
    \label{fig:attn_sparsity}
\end{figure}

\noindent \textbf{The attention can be highly sparse.}
We observe that only a small fraction of tokens in the attention mechanism are truly important.
We evaluate 128 randomly sampled data points from the WikiText-2 corpus~\cite{merity2016pointer}, analyzing two randomly selected attention heads from Qwen2-0.5B and Qwen2-1.5B. The results are shown in Figure~\ref{fig:attn_sparsity}.
For Qwen2-0.5B, more than 80\% attention scores fall below 0.01; over 90\% attention scores fall below 0.03.
A similar trend is observed in Qwen2-1.5B. These near-zero values contribute negligibly to the attention output and, therefore, need not be computed. In other words, the reliance of attention on CPU/GPU computation can be substantially reduced.

\begin{figure}[t]
    \centering
    \includegraphics[width=0.45\textwidth]{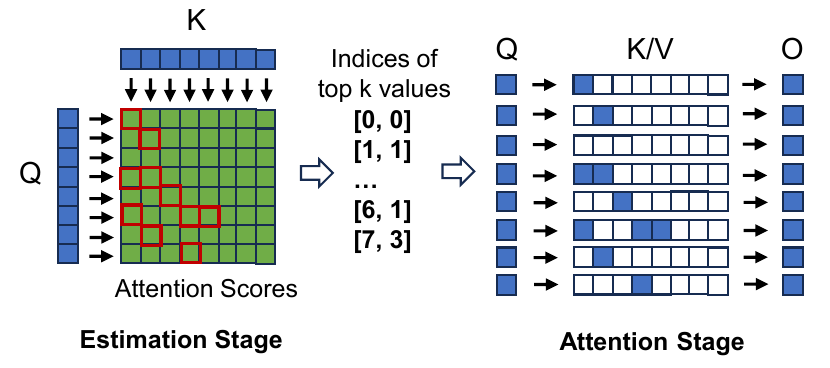}
    \caption{The workflow of sparse attention.}
    \label{fig:sparse_attn}
\end{figure}

\noindent \textbf{The workflow of sparse attention.}
Figure~\ref{fig:sparse_attn} shows the general workflow of a sparse attention.
This workflow is flexible enough to encompass various empirical sparsity patterns—such as sliding windows or vertical-slash patterns~\cite{lai2025flexprefillcontextawaresparseattention, jiang2023mistral7b, jiang2024minference10acceleratingprefilling}, and can also handle sparsity without obvious structural patterns.
It consists of two stages: (1) \textit{Estimation stage}.
An attention score matrix is calculated by $attentionScores=softmax(mask(\frac{ Q\cdot K}{\sqrt{d_{k}}}))$, with all tokens in Q and K involved.
After that, top k values out of the attention scores are selected, with their indices transferred to the next stage.
(2) \textit{Attention stage}.
Only the tokens that are retained by the indices from the estimation stage is involved in the sparse compute of $O=softmax(mask(\frac{ Q\cdot K}{\sqrt{d_{k}}}))\cdot V$.

\begin{figure}[t]
    \centering
      \begin{minipage}[b]{0.23\textwidth}
        \includegraphics[width=\textwidth]{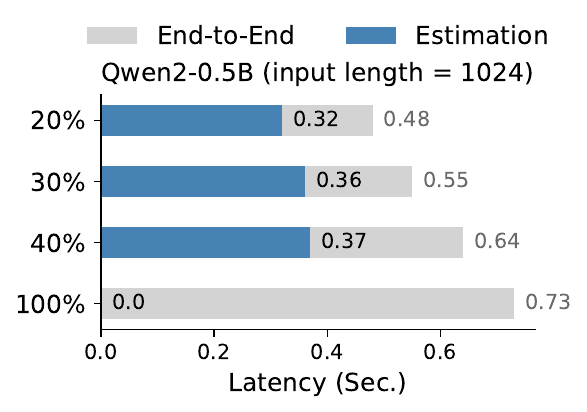}
        \subcaption{Latency breakdown.}
        \label{fig:breakdown_latency}
      \end{minipage}
      % \hfill
      \begin{minipage}[b]{0.24\textwidth}
        \includegraphics[width=\textwidth]{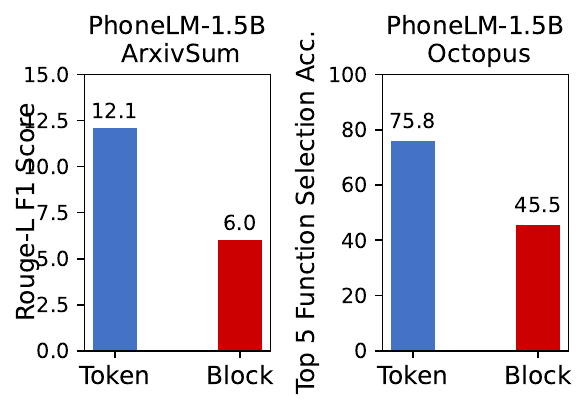}
        \subcaption{Accuracy drop.}
        \label{fig:block_acc_drop}
      \end{minipage}
    \caption{(a): The breakdown latency of an attention kernel that runs on a CPU mid core of snapdragon 8egn3 SoC. (b): The accuracy drop when grouping 64 tokens as a block for estimation on mobile LLMs and tasks.}
    \label{fig:estimation_overhead}
\end{figure}

\noindent \textbf{Bottleneck: the estimation stage.}
Despite the potential advantages of sparse attention, directly applying it in LLM inference provides little performance benefit.
% The key bottleneck of the sparse attention workflow lies in the estimation stage.
As shown in Figure~\ref{fig:breakdown_latency}, given a sparsity ratio of 40\%/30\%/20\%, the end-to-end latency of an attention kernel only speeds up by 12.4\%/24.7\%/34.3\%.
This is because, unlike the attention stage that computes only the important tokens, the estimation stage requires processing all tokens, dominating the overall computation when the attention is extremely sparse.
For instance, in a kernel with 20\% sparsity (80\% tokens are discarded), 66.7\% out of the end-to-end computation is attributed to the estimation.
% This is because that the estimation stage dominates when the compute is extremely sparse.
% For instance, in a kernel with 20\% sparsity (80\% tokens are discarded), 66.7\% out of the end-to-end latency is attributed to the estimation, since this stage involves all of the tokens.
An intuitive approach is to group the adjacent tokens in Q and K into a block before the estimation, and use the attention score of the entire block to represent the importance of each token in this block, i.e., block sparse attention~\cite{yang2025lserveefficientlongsequencellm}, reducing the estimation overhead.
However, the block sparse attention degrades the accuracy of mobile LLMs significantly, as it overlooks some important tokens due to its coarse-grained estimation.
For instance, in Figure~\ref{fig:block_acc_drop}, PhoneLM-1.5B shows a 6.1 pp accuracy drop on ArxivSum dataset.
\textit{
In a nut shell, \sys must carefully design the estimation stage to ensure high accuracy while keeping latency low.}

% \subsection{Challenges}

% \noindent \textbf{Hardware stack.}
% Int Dense Units (IDUs).
% Designed for low-precision, high throughput operations.
% Not friendly to high-precision, sparse operations.

% \noindent \textbf{Software stack.}
% Static graph.
% Not friendly to quantization (per-tensor, fixed scale).
% Less flexibility on dynamic shape.
% Coarse-grained control, black box, hard to program and debug.

% \noindent \textbf{Limitations to attention.}
% For dense attention, not accurate due to quantization, especially the per-tensor fixed scale quantization.
% For sparse attention, not accurate and not efficient.

% \subsection{Hardware Resource Abstraction}
% Propose introducing a Float Sparse Unit (FSU) to perform accurate and efficient sparse attention.

% Software stack is CPU-like, supports fine-grained control of dynamic graph.

% \subsection{Challenges}

% \textbf{The sparse ratio?}

% \noindent \textbf{The estimation overhead.}

% \noindent \textbf{The static graph on QxK.}

% \noindent \textbf{The idle of IDU-FSU.}

% At three levels

% \noindent \textbf{Run what.}

% \noindent \textbf{on which processor.}

% \noindent \textbf{how to schedule.}
\section{The Design}
\label{sec:design}

\subsection{Overview}
\label{sec:overview}

\begin{figure}[t]
    \centering
    \includegraphics[width=0.48\textwidth]{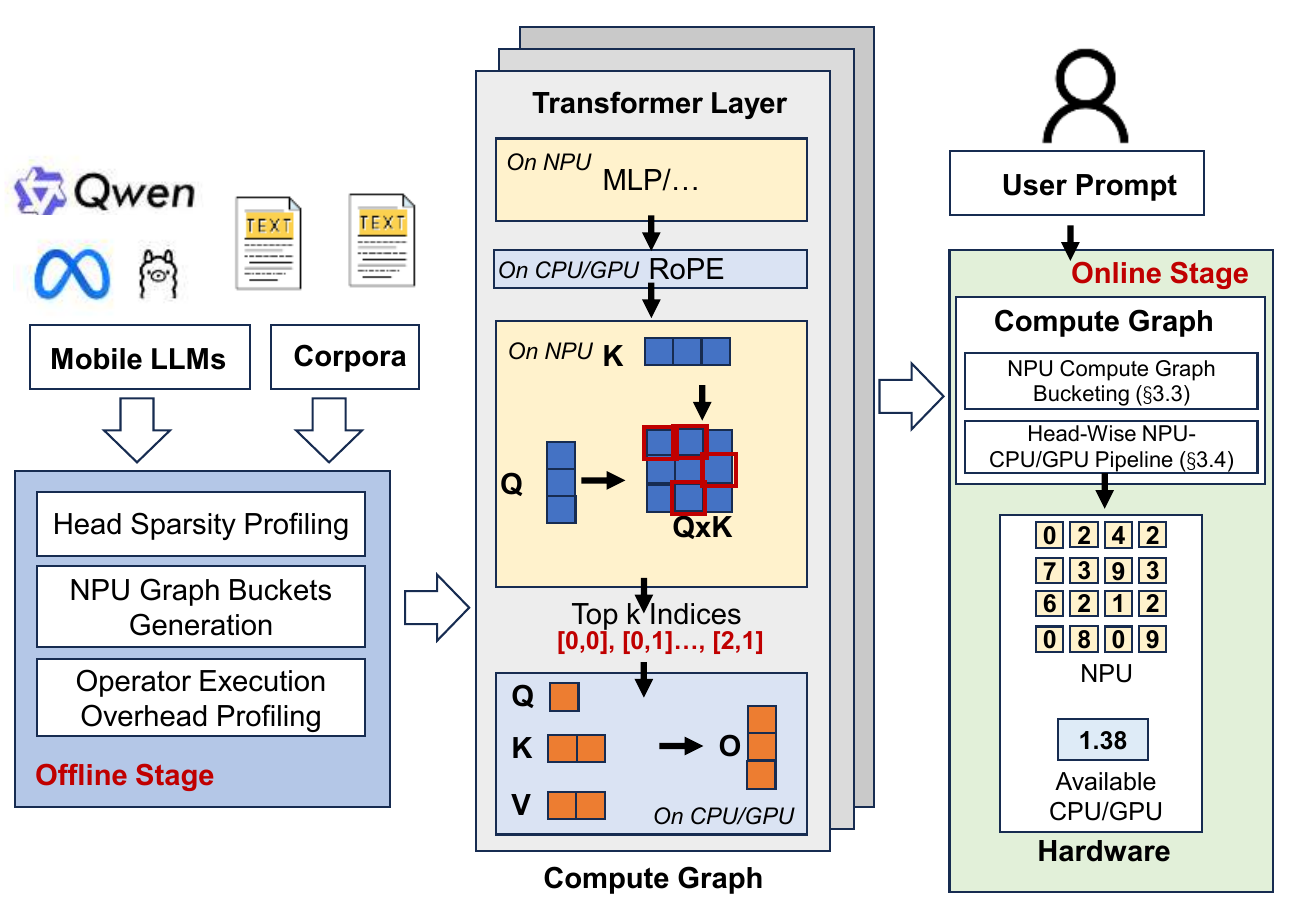}
    \caption{The workflow of \sys.}
    \label{fig:sys_overview}
\end{figure}

\noindent \textbf{Design Goal.}
\sys is a system–algorithm co-designed, NPU-centric LLM inference framework tailored for mobile SoCs. Its goal is to execute attention both accurately and efficiently, while minimizing reliance on general-purpose processors such as CPUs and GPUs.

\noindent \textbf{Workflow of \sys.}
Figure~\ref{fig:sys_overview} shows the workflow of \sys that consists the following parts.

\noindent $\bullet$ \textit{The offline stage.}
\sys performs several kinds of lightweight profiling on general corpora (by default 128 samples from WikiText-2).
Firstly, \sys determines a sparse ratio for each separated head based on its importance and  global sparsity ratio (configured by system).
% Their average value equals to a global sparsity ratio (configured by system).
Then, \sys generates multiple static graphs and organizes them in buckets for online inference.
Besides, it also profiles the on-device execution overhead of key operators on the SoC with OS-configured available CPU/GPU resource for pipeline planning.
% The motivation and technical details of them can be found in the following subsections.
This stage is mainly performed on cloud servers before installing the model to mobile devices, with negligible time and resource overhead (e.g., < 1 cloud GPU hours for each model).

\noindent $\bullet$ \textit{The NPU-centric compute graph.}
\sys generates a sparse attention module.
This module can be further integrated into the on-device inference frameworks as an end-to-end NPU-centric compute graph. 
As shown in Figure~\ref{fig:sys_overview}, the operators excluding the attention run on NPU by the inference framework. 
For the attention module, the RoPE is run on CPU/GPU, and the estimation of Q$\cdot$K is run on NPU in INT8.
After getting the top k positions, the CPU/GPU runs the fragmented QKV sparsely to get the output O.
The compute and data movement outside the NPU is minimized.
% By design, the sparse attention of \sys is also compatible with block sparse attention (tiny block size recommended, e.g., 2x2).
% In this case, only a pooling operator is needed to insert after the Q and K tensors.
% We elaborate the details of \sys's sparse attention in $\S$\ref{subsec:sparse_attn}.

% \xdl{
% Figure~\ref{fig:sys_overview} illustrates an end-to-end NPU-centric compute graph of \sys. 
% For the operators excluding the attention, they run on NPU.
% For the attention operators, \sys generates a sparse attention kernel for the limited CPU/GPU resource, where the RoPE is run on CPU/GPU, and the estimation of Q$\cdot$K is run on NPU in INT8.
% After getting the top k positions, the CPU/GPU runs the fragmented QKV sparsely to get the output O.
% The compute and data movement outside the NPU is minimized.
% }

\noindent $\bullet$ \textit{The online stage.}
The compute graph takes the user's prompt text as the input, and executes each part on the CPU/GPU/NPU of mobile SoC.
During inference, two optimizations are further introduced, i.e., NPU compute graph bucketing  ($\S$\ref{subsec:buckets}) and Head-wise NPU-CPU/GPU pipeline ($\S$\ref{subsec:pipeline}).

% \noindent \textbf{Key techniques.}

\subsection{Dynamic Sparse Attention of \sys}
\label{subsec:sparse_attn}

\begin{figure}[t]
    \centering
      \begin{minipage}[b]{0.21\textwidth}
        \includegraphics[width=\textwidth]{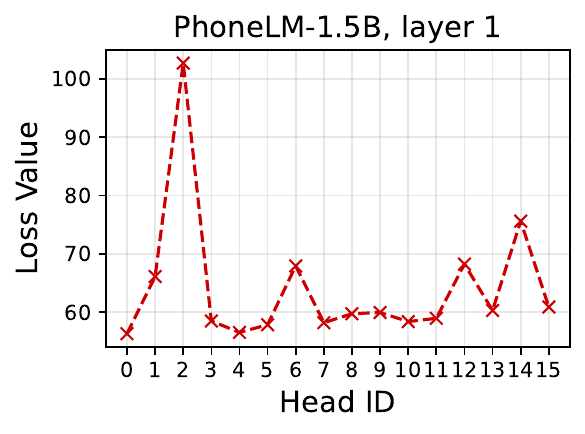}
        \subcaption{Loss on each head.}
        \label{fig:head_imp}
      \end{minipage}
      % \hfill
      \begin{minipage}[b]{0.26\textwidth}
        \includegraphics[width=\textwidth]{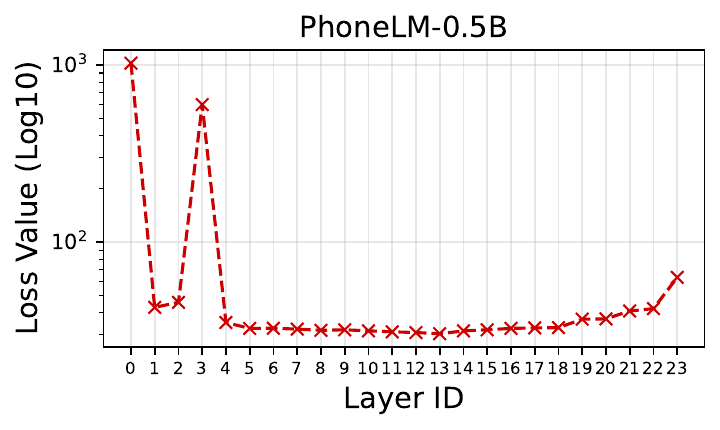}
        \subcaption{Loss on each layer.}
        \label{fig:layer_imp}
      \end{minipage}
    \caption{The importance is uneven across heads and layers.(a): Removing the heads in layer 1 of PhoneLM-0.5B; (b) removing the layers of PhoneLM-0.5B. The data is on 128 samples of WikiText-2. Loss values over 1e-3 are clamped to 1e-3. The y-axis of subfigure (b) is processed by log10.}
    \label{fig:head_layer_imp}
\end{figure}

% \usepackage{multirow}
\begin{table}[]
\small
\begin{tabular}{c|ccccc}
\hline
\multirow{2}{*}{\textbf{Models}} & \multicolumn{5}{c}{\textbf{Various global sparse ratios}}                     \\ \cline{2-6} 
                                 & \textbf{20\%} & \textbf{30\%} & \textbf{40\%} & \textbf{50\%} & \textbf{80\%} \\ \hline
PhoneLM-0.5B                     & 99.32\%       & 99.39\%       & 99.45\%       & 99.51\%       & 99.79\%       \\
PhoneLM-1.5B                     & 99.24\%       & 99.27\%       & 99.29\%       & 99.34\%       & 99.61\%       \\
Qwen2-0.5B                       & 99.03\%       & 99.32\%       & 99.51\%       & 99.59\%       & 99.86\%       \\
Qwen2-1.5B                       & 99.68\%       & 99.80\%       & 99.83\%       & 99.84\%       & 99.95\%       \\ \hline
\end{tabular}
\caption{Predicting the important positions in the Q$\cdot$K via NPU. Data: 128 samples from WikiText-2. We report the average recall rate of all heads with the float QxK as ground truth. The rate is surprisingly high under various global sparse ratios.}
\label{tab:qxk_recall}
\end{table}

\noindent \textbf{Head-specific sparse ratio.}
One of \sys's insight is that the sparse ratio of attention should be head-specific.

According to the progress of eXplainable AI (XAI)~\cite{Mersha_2024}, the importance of a certain module of a neuron network can be measured by the delta loss after removing it.
For instance, on a calibration dataset $C$, the importance of head $i$ in layer $j$ can be measured as
\begin{equation}
\begin{split}
    headImp_{i} = loss(head_{i}=0, C) - loss(C),
\end{split}
\label{formula:head_importance}
\end{equation}
and the layer j where the head is located can be measured by
\begin{equation}
\begin{split}
    layerImp_{j} = loss(layer_{j}=0, C) - loss(C).
\end{split}
\label{formula:head_importance}
\end{equation}

In Figure~\ref{fig:head_layer_imp}, we show the importance of head $i$ and layer $j$.
We remove the heads in the first layer in Figure~\ref{fig:head_layer_imp}(a), and show the loss after removal.
The results exhibit obvious unevenness.
For instance, the head 2 makes the loss value increase to over 100, yet the head 8 only makes the loss value increase to less than 60.
Such an observation can also be found in Figure~\ref{fig:head_layer_imp}(b), which depicts the layer-level loss.
The rationale behind the unevenness is mainly that various parts of a neuron network may learn various aspects of the data~\cite{Cai_2025, jiang2024minference10acceleratingprefilling, ge2024modeltellsdiscardadaptive}, especially in compact mobile LLMs.

In summary, for different heads in different layers, the sparsity ratio should also be uneven.
An important head should retain more tokens, while a trivial head should retain less.
Given a global average sparsity ratio $r$ and the number $N$ of all heads in a model, the sparsity ratio of $head_{i}$ is
\begin{equation}
\begin{split}
     \frac{r\cdot N\cdot clamp(headImp_{i}\cdot layerImp_{j})}{sum(clamp(headImp_{i}\cdot layerImp_{j}))}, 1 \leq i \leq N, 
\end{split}
\label{formula:sparsity_ratio}
\end{equation}
where $clamp(\cdot)$ means clamping extremely large points (e.g., over 1e-3) and $j=i/headNumPerLayer$.
In doing so, the intrinsic feature of attention heads is fully explored.

The sparse ratio of each head is determined at offline stage and stays fixed during online stage, incurring no overhead for inference.
The default calibration dataset is 128 samples from WikiText-2.
The offline profiling only takes about 5 minutes for a mobile LLM on a cloud server with a single A100 GPU, being affordable for most  developers.

\noindent \textbf{Running estimation on NPU.}
Another key insight of \sys is that the estimation can be offloaded to low-precision NPU.
% Due to the limitations of static graph and per-tensor quantization, the NPU is not friendly to attention.
% On the other hand, running the dense QxK estimation on the CPU/GPU leads to heavy overhead.
\sys's observation is that only determining the important positions in Q$\cdot$K is less prone to quantization compared to calculating the exact value of the QKV result.
For instance, Table~\ref{tab:qxk_recall} shows the average recall rate of predicting the important positions in the Q$\cdot$K via NPU's INT8 per-tensor quantization.
Surprisingly, the rate is more than 99\% for various models and global sparsity ratios.
This stands in stark contrast to the significant accuracy drop observed (e.g., 40.5\% on Octopus, PhoneLM-0.5B) when using NPU to compute QKV.
The rationale behind it is not hard to comprehend.
The computation of QKV results requires precise ``value equality'', whereas identifying the positions of larger elements in Q$\cdot$K allows the computational results to fluctuate within a certain range.
For example, the latter accepts both a Q$\cdot$K = \{0.4, 0.3, 0.2, 0.1\} and a quantized Q$\cdot$K = \{0.5, 0.4, 0.08, 0.02\}, while the former only accepts the non-quantized Q$\cdot$K. 

On top of this, \sys's workflow of estimation stage is as follows.
Firstly, it quantizes the Q and K tensor and runs dense INT8 Q$\cdot$K on NPU.
Then, the Q$\cdot$K result is transferred to CPU/GPU to find the top k values.
Finally, the CPU/GPU computes the sparse QKV according to the positions of the top k values.
A detail here is that \sys does not further compute the masked attention score based on Q$\cdot$K.
Since the softmax operation is strictly monotonically increasing, we can directly use the Q$\cdot$K that before softmax for top k.
Also, the causal mask is not applied to Q$\cdot$K.
Instead \sys straightforwardly skips the masked positions for the top k operation.
The skip is very convenient on CPU/GPU.

\subsection{NPU Compute Graph Bucketing}
\label{subsec:buckets}

\begin{figure}[t]
    \centering
    \includegraphics[width=0.48\textwidth]{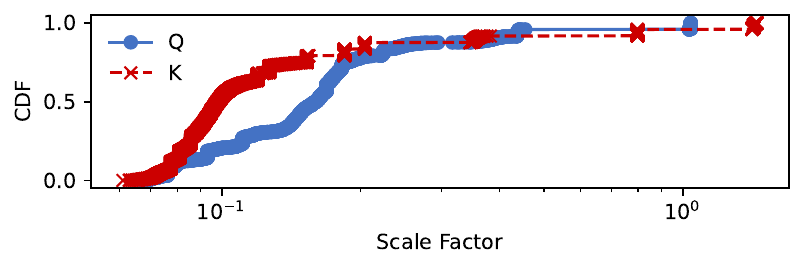}
    \caption{The CDF of each head's scale factors of Q/K. Model: Qwen2-0.5B; data: 128 samples from WikiText-2. The x axis is logged by 10.}
    \label{fig:scales_cdf}
\end{figure}
\begin{figure}[t]
    \centering
    \includegraphics[width=0.48\textwidth]{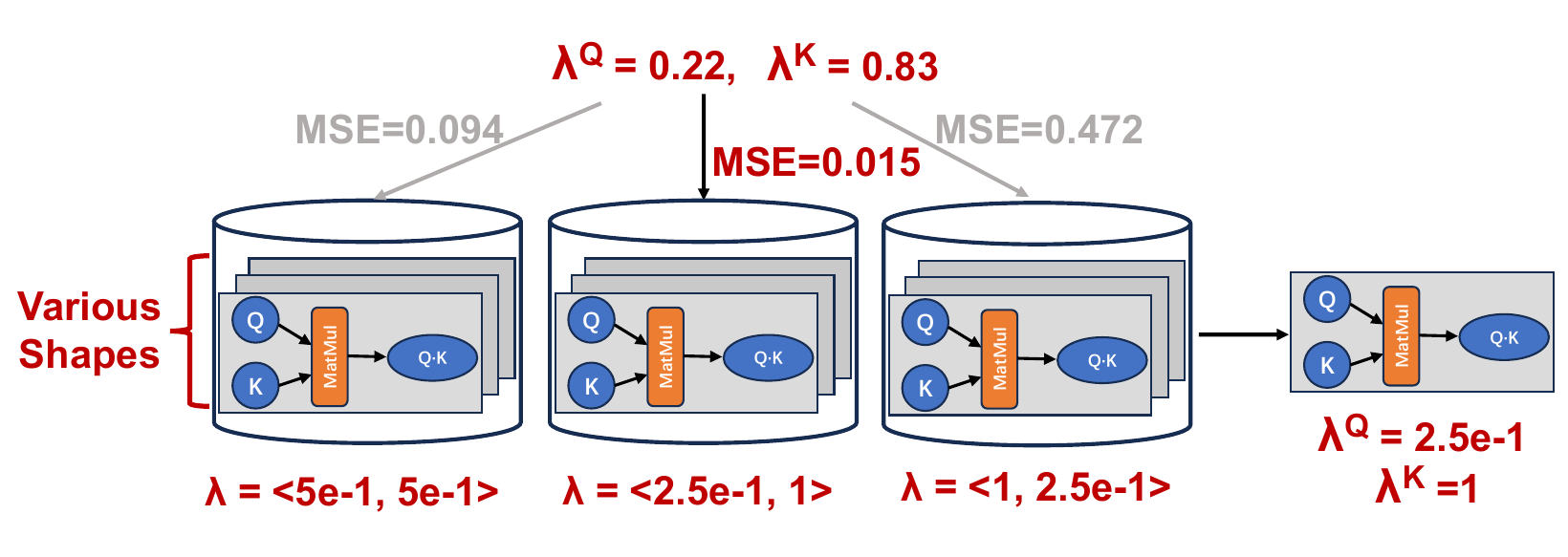}
    \caption{Bucketing the graphs to handle varying inputs.}
    \label{fig:scale_buckets}
\end{figure}

\noindent \textbf{The fluctuation of scale factors.}
As shown in Figure~\ref{fig:scales_cdf}, we show the scale factors of each head in Q tensor and K tensor for NPU-based estimation stage, with a per-tensor (per-head) linear quantization of INT8.
The data samples are the same to the aforementioned.
The results show that the value of the scale factor varies significantly for each head in both Q and K.
For instance, over 30\% scale factors in Q tensor are larger than 0.2; in comparison, 20\% scale factors are smaller than 0.1.
Recall that to maximize the runtime efficiency, the software stack of NPU requires offline compiling a static compute graph that with fixed constant and shape.
If we apply a single graph to all inputs, the accuracy will drop significantly since there is no one-size-fits-all scale factor as the graph constant.

% Besides, the graph shape can be also dynamic.
% On mobile devices, there is almost no batched input, and the sequence dimension is also fixed thanks to the chunked inference feature in modern on-device Transformer inference frameworks.
% However, there can be 

\noindent \textbf{Our solution: bucketing the graphs.}
\sys offline generates multiple graphs and cache them in buckets for online usage.
Specifically, as shown in Figure~\ref{fig:buckets}, we denote $\lambda^{Q}$/$\lambda^{K}$ as the scale factor of Q/K, and $\lambda=<\lambda^{Q}$, $\lambda^{K}>$ as the constant of a QxK compute graph.

At offline stage, we profile the average $\Bar{\lambda}$ on the calibration corpora, and generate $\{<\Bar{\lambda}^{Q}, \Bar{\lambda}^{K}>, <\Bar{\lambda}^{Q}\cdot\sigma, \Bar{\lambda}^{K}/\sigma>, \cdots, <\Bar{\lambda}^{Q}\cdot\sigma, \Bar{\lambda}^{K}\cdot\sigma>\}$ buckets, where $\sigma$ is the step size (by default 5e-1, detailed later in $\S$\ref{subsec:sensitivity_ablation}).
Notably, a bucket contains multiple graphs with various shapes\footnote{The ``shape'' here mainly refers to the head dimension. On mobile devices, there is almost no batched input, and the sequence dimension is also fixed thanks to the chunked inference feature in modern on-device Transformer inference frameworks~\cite{llm.npu}.} due to the NPU kernel fused launch (detailed in $\S$\ref{subsec:pipeline}).
The memory overhead of caching the graphs is negligible.
Since a QxK graph does not contain any weight, a graph only takes about 10--100 KB.

At online stage, in Figure~\ref{fig:buckets}, given a pair of input tensors Q and K, we first calculate the MSE of $<\lambda^{Q}, \lambda^{K}>$ to each bucket's $\lambda$.
Then, the input falls into the bucket with the smallest MSE.
Finally, a compute graph with the corresponding shape is selected out for the following inference.
In doing so, the accuracy of NPU estimation is guaranteed.
The configuration details of bucketing the graphs are discussed in $\S$\ref{sec:eval}. 

\subsection{Head-Wise NPU-CPU/GPU Pipeline}
\label{subsec:pipeline}

\begin{figure*}[t]
    \centering
    \includegraphics[width=0.99\textwidth]{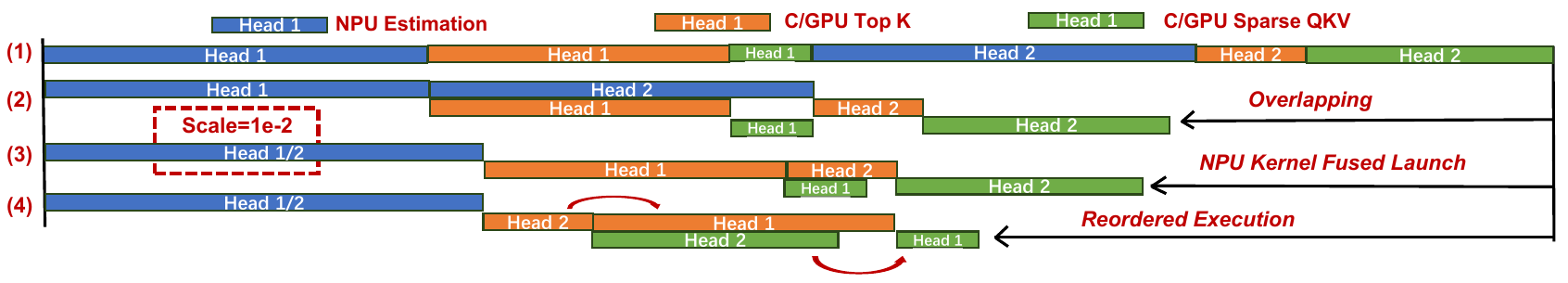}
    \caption{An illustration of NPU-CPU/GPU pipeline.}
    \label{fig:IDUFSUPipeline}
\end{figure*}

Consider the execution of \sys's dynamic sparse attention.
Figure~\ref{fig:IDUFSUPipeline} shows a minimal illustration with two heads.
In each head, we need to first estimate the attention scores (i.e., the blue section), then select top k positions (i.e., the orange section), and finally do sparse QKV operations.
Formally, if we use ${\zeta}^{i}_{npu}$, ${\zeta}^{i}_{topk}$, and ${\zeta}^{i}_{qkv}$ to represent executing the corresponding section of the $i_{th}$ of $n$ heads, the execution obeys
\begin{equation}
\begin{split}
    {\zeta}^{i}_{npu} \leftarrow {\zeta}^{i}_{topk}; \quad {\zeta}^{i}_{npu}, {\zeta}^{i}_{topk} \leftarrow {\zeta}^{i}_{qkv}, \forall i \in n,
\end{split}
\label{formula:dependency}
\end{equation}
where ``$\leftarrow$'' means the dependency.

The naive way is running each operation sequentially (Figure~\ref{fig:IDUFSUPipeline}(1)).
However, this ignores several key optimizations in this procedure.
\sys further introduces the following insights.

\noindent \textbf{Overlapping.}
Both ${\zeta}^{i}_{npu}$ and ${\zeta}^{i}_{qkv}$ operations are compute-bound, and the ${\zeta}^{i}_{topk}$ is relatively memory-bound.
Furthermore, ${\zeta}^{i}_{npu}$ and ${\zeta}^{i}_{qkv}$ run on separate processors (NPU and CPU/GPU, respectively).
Thus, according to Formula~\ref{formula:dependency}, if $ {\zeta}^{i}_{npu}$ $ {\zeta}^{j}_{topk}$ and $ {\zeta}^{k}_{qkv}$ have no dependency, then they can be executed in parallel. 
As shown in Figure~\ref{fig:IDUFSUPipeline}(2), after overlapping these operations by a three-stage pipeline, the end-to-end latency can be reduced. 

\noindent \textbf{NPU kernel fused launch.}
The NPU is designed for high-throughput dense operations.
A vanilla method of launching a single head for execution causes under-utilization of the IDU.
For instance, the QxK compute graph of Q=[1,1,128,64] and K=[1,1,2048,64] (layouted in BHSD) takes 2ms on MI14; a graph that fuses 2 heads or 4 heads together only takes 3 ms or 4ms, respectively.
On the other hand, since the number of scale factor buckets is finite (in fact a few buckets are enough, as detailed in $\S$\ref{subsec:sensitivity_ablation}), there exist many heads share the same scale factor.
Thus, such heads can be fused together in a launch.
As shown in Figure~\ref{fig:IDUFSUPipeline}(3), head 1 and head 2 have the same scale factor.
After fusing, the pipeline latency is further reduced.

\noindent \textbf{Reordered execution.}
There exist bubbles in the pipeline due to the dependency in Formula~\ref{formula:dependency}.
Since each head has its own sparsity ratio, the execution order of the heads influences the bubbles.
A good order can squeeze out more bubbles.
As shown in Figure~\ref{fig:IDUFSUPipeline}(4), executing head 2 before head 1 further achieves latency reduction.

\noindent \textbf{The pipeline planning policy.}
\sys's pipeline puts the above optimizations together.
Yet, solving the planning of such a pipeline has O(n!) time complexity, which is NP-hard and unacceptable on mobile devices.
To this end, \sys employs a greedy search with polynomial complexity in Algorithm~\ref{algo:pipeline}.
Basically, we first do fused launch at the beginning of planning (line 15).
Then, for each planning step, we select the ${\zeta}^{i}_{npu}$, ${\zeta}^{i}_{topk}$ and ${\zeta}^{i}_{qkv}$ that minimize the current pipeline latency.
The overhead of ${\zeta}^{i}_{npu}$, ${\zeta}^{i}_{topk}$ and ${\zeta}^{i}_{qkv}$ is obtained in the lightweight offline stage mentioned in $\S$\ref{sec:overview}.
Such a pipeline planning policy incurs negligible online overhead on mobile devices (e.g. < 1 ms on MI14).

\begin{algorithm}[t]
\footnotesize
\caption{The pipeline planning policy.}\label{planning}

\noindent\textbf{Input: } $heads_{c/gpu}$: each head of the attention.
\noindent\textbf{Output: } $res_{npu}$, $res_{c/gpu}$: the execution order. $t_{qkv}$: the total execution time.
\begin{algorithmic}[1]
\State \textit{\textbf{C/GPUPlan}}($t_{npu}$, $t_{topk}$, $t_{qkv}$, $fusedHeads$)
    \State \quad $res$ = []
    \State \quad \textbf{for} $head$ in $fusedHeads$
        \State \quad\quad $t_{min}$ = INF, $selectedHead$ = -1
        \State \quad\quad \textbf{for} $head$ in $fusedHeads$
            \State \quad\quad\quad \textbf{if} $head$ in $res$ \textbf{then} continue
            \State \quad\quad\quad $t^{new}_{topk}$ = max($t_{npu}$, $t_{topk}$) + topk time of $head$
            \State \quad\quad\quad $t^{new}_{qkv}$ = max($t_{qkv}$, $t_{topk}^{new}$) + qkv time of $head$
            \State \quad\quad\quad \textbf{if} $t_{qkv}^{new}$ < $t_{min}$ \textbf{then} $t_{min}$ = $t_{qkv}^{new}$, $selectedHead$ = $head$
        \State \quad\quad $res$.append($selectedHead$)
        \State \quad\quad $t_{topk}$ = max($t_{npu}$, $t_{topk}$) + topk time of $selectedHead$
        \State \quad\quad $t_{qkv}$ = max($t_{qkv}$, $t_{topk}$) + qkv time of $selectedHead$
    \State \quad \textbf{return} $res$, $t_{topk}$, $t_{qkv}$
\State \textit{\textbf{Plan}}($heads_{c/gpu}$)
    \State \quad $heads_{npu}$ $\gets$ $heads_{c/gpu}$ \Comment{fuse the heads}
    \State \quad $t_{npu}$ = 0, $t_{topk}$ = 0, $t_{qkv}$ = 0, $res_{npu}$ = [], $res_{c/gpu}$ = []
    \State \quad \textbf{for} $fusedHeads$ in $heads_{npu}$ \Comment{a planning step}
        \State \quad\quad $t_{min}$ = INF, $selectedHeads$ = None, $res$ = []
        \State \quad\quad \textbf{for} $fusedHeads$ in $heads_{npu}$ \Comment{greedy search}
            \State \quad\quad\quad \textbf{if} $fusedHeads$ in $res_{npu}$ \textbf{then} continue
            \State \quad\quad\quad $t_{npu}^{new}$ = $t_{npu}$ + npu time of $fusedHeads$
            \State \quad\quad\quad $res$, $t_{topk}^{new}$, $t_{qkv}^{new}$ = \textbf{\textit{C/GPUPlan}}($t_{npu}^{new}$,$t_{topk}$,$t_{qkv}$,$fusedHeads$)
            \State \quad\quad\quad \textbf{if} $t_{qkv}^{new}$ < $t_{min}$ \textbf{then} 
                \State \quad\quad\quad\quad $t_{min}$ = $t_{qkv}^{new}$, $selectedHeads$ = $fusedHeads$
        \State \quad\quad $res_{npu}$.append($selectedHeads$), $res_{fsu}$ += $res$
        \State \quad\quad $t_{npu}$ = $t_{npu}$ + npu time of $selectedHeads$
        \State \quad\quad $\_$, $t_{topk}$, $t_{qkv}$ = \textbf{\textit{C/GPUPlan}}($t_{npu}$, $t_{topk}$, $t_{qkv}$, $selectedHeads$)
    \State \quad \textbf{return} $res_{npu}$, $res_{c/gpu}$, $t_{qkv}$
\end{algorithmic}
\label{algo:pipeline}
\end{algorithm}
\section{Implementation}
\label{sec:impl}

We prototype \sys with ~10,000 LoC in C++/python.
We choose COTS smartphones with Qualcomm SoCs as the testbed, and \sys can also work on other mainstream mobile devices.
We only apply a middle core of ARM CPU as the minimal available CPU/GPU resource.
We vary the available resource in the following experiments in $\S$\ref{subsec:sensitivity_ablation}.
% Only using a single non-prime core still aligns well with the ``accelerator-only'' idea of \sys.
The NPU's software stack is built upon QNN~\cite{qnn} and Hexagon SDK~\cite{hexagon_sdk} of Snapdragon NPU.
The CPU/GPU's software stack consists of ARM NEON~\cite{arm_neon}, OpenCL~\cite{opencl} and LLVM ecosystem~\cite{llvm}, preserving the most runtime flexibility like dynamic graph execution.
We fuse several operations, such as RoPE~\cite{su2023roformerenhancedtransformerrotary}, quantization, matmul and softmax together in the kernel to minimize the redundant memory copy. 
Some non-sensitive instructions like multiplication inside them are further executed in half precision.

\sys is compatible with any on-device LLM inference library.
It can be integrated by directly including a header file, being least intrusive to the original codebase.
In the following end-to-end experiments, we choose a high performance framework llm.npu~\cite{llm.npu}, whose non-attention parts can run on NPU.
To the best of our knowledge, this is the only fully open-source framework running on NPUs.
Specifically, we replace the attention operator with \sys and its baselines for prefilling stage, and directly employ full attention on CPU/GPU for decoding since this stage is mainly memory-bound.
\section{Evaluation}
\label{sec:eval}

\noindent \textbf{Devices.}
We mainly perform the experiments on MI14~\cite{mi14} and Redmi K60 Champion Edition~\cite{k60}.
MI14 is equipped with Qualcomm 8Gen3 SoC and 16+6GB DRAM; Redmi K60 Champion Edition has 8Gen2 and 16+3GB DRAM.
The available NPU and CPU/GPU resources are Hexagon-V75/V73 and a mid-core Cortex-A720/A715 processor, respectively.

\noindent \textbf{Models.}
We mainly test \sys on mobile LLMs.
\textit{(1) Qwen2-0.5B/1.5B}~\cite{qwen2-0.5b, qwen2-1.5b}; \textit{(2) PhoneLM-0.5B/1.5B}~\cite{phonelm-0.5b, phonelm-1.5b}.
The details are listed in Table~\ref{tab:models}.
The model weights are quantized by third-party frameworks in the following end-to-end experiments.

% \usepackage{booktabs}
\begin{table}[]
\small
\begin{tabular}{@{}ccccc@{}}
\toprule
Model        & Q Heads & KV Heads & Dims & Layers \\ \midrule
PhoneLM-0.5B~\cite{phonelm-0.5b} & 16      & 16       & 64   & 24     \\
PhoneLM-1.5B~\cite{phonelm-1.5b} & 16      & 16       & 160  & 19     \\
Qwen2-0.5B~\cite{qwen2-0.5b}   & 14      & 2        & 64   & 24     \\
Qwen2-1.5B~\cite{qwen2-1.5b}   & 12      & 2        & 128  & 28     \\ \bottomrule
\end{tabular}
\caption{Models we use in our experiments.}
\label{tab:models}
\end{table}

\noindent \textbf{Datasets.}
We test \sys on the following datasets.

\textit{(1) ArxivSum}~\cite{cohan-etal-2018-discourse} is a general natural language comprehension task. 
We report the Rouge-L F1 score.

\textit{(2) DroidCall}~\cite{xie2024droidcalldatasetllmpoweredandroid} is an LLM agent task on mobile GUI automation.
We report the top 5 single step complete accuracy.

\textit{(3) Octopus}~\cite{chen2024octopusv2ondevicelanguage} is an LLM agent task on mobile system API function calling.
We report the top 5 function selection accuracy.

Each test data is augmented by in-context learning with 5 samples.

\noindent \textbf{Baselines.} We compare \sys to the following design alternatives, with fully available NPU and highly limited CPU/GPU resource, i.e., one middle core of CPU.

\textit{(1) Full Attention on CPU/GPU (C/G-Full)} runs float32 full atention on CPU/GPU, a default setting of mainstream on-device inference frameworks, such as llm.npu~\cite{llm.npu}, mlc-llm~\cite{mlc-llm} and HeteroLLM~\cite{chen2025heterollmacceleratinglargelanguage}.

\textit{(2) Sparse Attention on CPU/GPU (C/G-Sparse)} runs float32 dynamic sparse attention on CPU/GPU, including the dense estimation procedure.
This is a vanilla setting of fine-grained dynamic sparse attention~\cite{wang2021spatten, zhang2023h2oheavyhitteroracleefficient}.

\textit{(3) Block Sparse Attention on CPU/GPU (C/G-Block-Sparse)} runs float32 dynamic sparse attention with a 64x64 block size, which is a common setting in previous literature~\cite{yang2025lserveefficientlongsequencellm}.  

\textit{(4) Full Attention on NPU (NPU-Full)} runs INT8 full attention on NPU, with static graph and per-tensor quantization~\cite{xue2024powerinfer2fastlargelanguage}.

\noindent \textbf{Settings.}
In the following experiments, we set global sparsity ratio as 20\%, and the number of scale buckets as 9 (step size 5e-1). 
The rationale of selecting them is detailed in $\S$\ref{subsec:sensitivity_ablation}.
The default available CPU/GPU resource for LLM inference is set to one mid core of CPU.

\subsection{End-to-End Performance}

\textbf{\sys is accurate.}
We first report \sys's accuracy on the aforementioned datasets and models against its baselines.
As shown in Table~\ref{tab:acc}, \sys is the most accurate alternative compared to the lossless baseline C/G-Full, with only 0.4 pp.
This is due to \sys's well-designed dynamic sparsity, specifically the fine-grained token-level estimation and head-specific sparsity.
Other baselines show significant accuracy loss compared to C/G-Full, with 7.4/11.4/18 pp.
The reasons are many fold.
The basic reason is that sparsity introduces more noise into the model, especially for compact mobile LLMs and datasets with high information density.
For C/G-Block-Sparse, the blocked sparsity makes the model more easier to ignore important tokens.
For NPU-Full, the limitations of static graphs and per-tensor quantization result in inaccuracy, particularly for the attention operator, where inputs are activations with high fluctuation.

% \usepackage{booktabs}
% \usepackage{multirow}
\begin{table}[]
\footnotesize
\begin{tabular}{@{}ccccccc@{}}
\toprule
\textbf{Model} &
  \textbf{Dataset} &
  \textbf{\begin{tabular}[c]{@{}c@{}}C/G-\\ Full\end{tabular}} &
  \textbf{\begin{tabular}[c]{@{}c@{}}C/G-\\ Sparse\end{tabular}} &
  \textbf{\begin{tabular}[c]{@{}c@{}}C/G-Block\\ -Sparse\end{tabular}} &
  \textbf{\begin{tabular}[c]{@{}c@{}}NPU-\\ Full\end{tabular}} &
  \textbf{Ours} \\ \midrule
\multirow{3}{*}{\begin{tabular}[c]{@{}c@{}}PhoneLM\\ -0.5B\end{tabular}} & ArxivSum  & 14.7 & 14.9 & 10.0 & 0.0  & 15.2 \\
                                                                         & DroidCall & 27.5 & 24.0 & 25.5 & 20.5 & 25.5 \\
                                                                         & Octopus   & 64.6 & 71.3 & 62.9 & 24.1 & 64.0 \\ \cmidrule(l){2-7} 
\multirow{3}{*}{\begin{tabular}[c]{@{}c@{}}PhoneLM\\ -1.5B\end{tabular}} & ArxivSum  & 11.9 & 12.1 & 6.0  & 0.0  & 12.0 \\
                                                                         & DroidCall & 20.5 & 23.5 & 28.5 & 19.0 & 25.5 \\
                                                                         & Octopus   & 79.2 & 75.8 & 45.5 & 24.7 & 75.8 \\ \cmidrule(l){2-7} 
\multirow{3}{*}{\begin{tabular}[c]{@{}c@{}}Qwen2\\ -0.5B\end{tabular}}   & ArxivSum  & 10.7 & 11.4 & 10.8 & 9.4  & 11.3 \\
                                                                         & DroidCall & 34.5 & 34.5 & 26.0 & 27.5 & 37.5 \\
                                                                         & Octopus   & 60.6 & 42.1 & 37.6 & 34.8 & 61.2 \\ \cmidrule(l){2-7} 
\multirow{3}{*}{\begin{tabular}[c]{@{}c@{}}Qwen2\\ -1.5B\end{tabular}}   & ArxivSum  & 8.5  & 2.0  & 4.2  & 9.1  & 8.4  \\
                                                                         & DroidCall & 48.0 & 19.0 & 15.5 & 22.5 & 44.5 \\
                                                                         & Octopus   & 61.2 & 21.9 & 32.0 & 34.2 & 56.1 \\ \midrule
\multicolumn{2}{c}{\textbf{Average}} &
  \textit{\textbf{36.8}} &
  \textit{\textbf{29.4}} &
  \textit{\textbf{25.4}} &
  \textit{\textbf{18.8}} &
  \textit{\textbf{36.4}} \\ \bottomrule
\end{tabular}
\caption{The accuracy on various datasets and models.}
\label{tab:acc}
\end{table}

\begin{figure*}[t]
    \centering
    \includegraphics[width=0.99\textwidth]{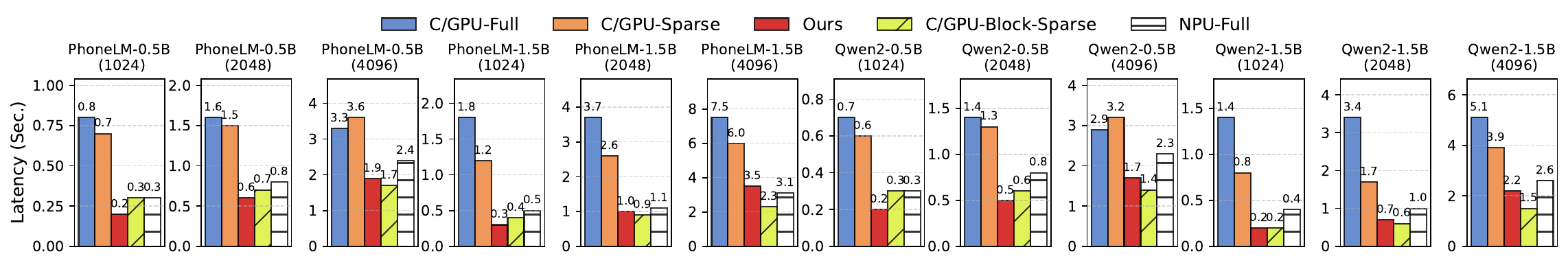}
    \caption{With the same circumstance of highly limited CPU/GPU resources, \sys can achieve much lower attention kernel latency compared to other baselines on MI14.}
    \label{fig:kernel_latency}
\end{figure*}
\begin{figure*}[t]
    \centering
    \includegraphics[width=0.99\textwidth]{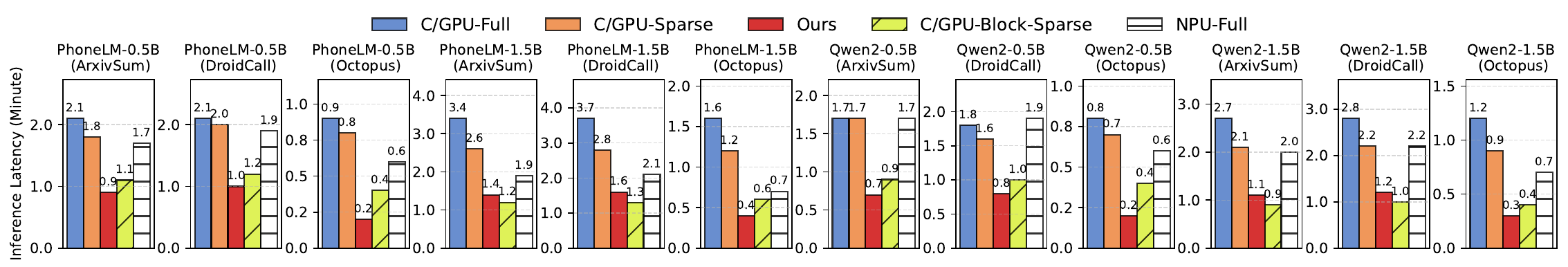}
    \caption{With the same circumstance of highly limited CPU/GPU resources, \sys can achieve much lower end-to-end average inference latency on datasets of real-world mobile tasks compared to other baselines on MI14.}
    \label{fig:e2e_latency}
\end{figure*}
\begin{figure*}[t]
    \centering
    \includegraphics[width=0.99\textwidth]{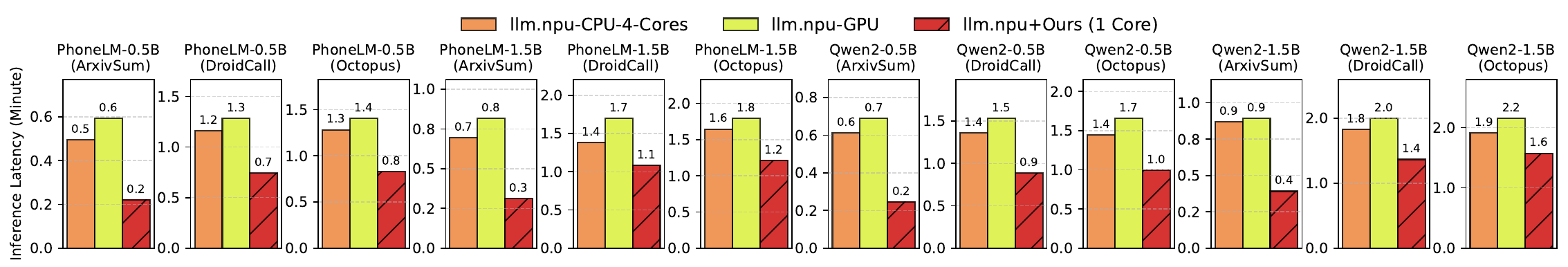}
    \caption{Compared to the native attention in SoTA NPU inference framework that shows high reliance on CPU/GPU, \sys achieves on-par or lower latency with significantly fewer CPU/GPU resources. Device: MI14.}
    \label{fig:e2e_latency_various_cpugpu}
\end{figure*}

\noindent \textbf{\sys is fast.} 
As shown in Figure~\ref{fig:kernel_latency}, we first report the latency of the attention kernel on MI14.
We test the kernel with various input length, i.e., 1024/2048/4096.
The values of the input tensors are randomized, and the sparisty ratio of each head is set to global ratio.
We have the following observations.
Compared to the lossless baseline C/G-Full, \sys is up to 6.9$\times$ and on average 3.5$\times$ faster.
Compared to the lossy baselines, \sys is up to 4.0$\times$ faster.
Note that according to Table~\ref{tab:acc}, employing these baselines for inference will result in significant accuracy loss.

We further report the end-to-end latency with integrating \sys and its baselines into third-party frameworks.
By doing so, we can realize NPU-centric inference.
The details are mentioned in $\S$\ref{sec:impl}.
We perform the experiments on three datasets, whose input sequence dominates the inference (3840/4096/1792 for prefill and 50/10/10 for decode, respectively).
This is the most common workload pattern on mobile devices.
The results are shown in Figure~\ref{fig:e2e_latency}.
Compared to the lossless baseline C/G-Full, \sys is up to 4.5$\times$ and on average 2.9$\times$ faster.
Compared to the lossy baselines, \sys is up to 4.0$\times$ faster.

We briefly discuss the rationales behind the experimental results.
For C/G-Sparse, although it enjoys a 20\% sparsity, the speedup is relatively low compared to C/G-Full.
This is due to the overhead of estimating the to-be-sparsified positions and the top-k operation.
For C/G-Block-Sparse, the speedup is higher.
This is due to the estimation overhead is lowered by pre-pooling the Q and K input tensors.
For NPU-Full, although NPU has high throughput on matrix multiplication, the speedup is somewhat reduced by operations such as softmax and masking.
In comparison, \sys achieves the highest speedup.
This is because that it minimizes the estimation overhead by offloading the estimation to NPU and overlapping NPU and CPU/GPU execution.

Besides, \sys achieves on-par or even lower performance compared to SoTA NPU framework llm.npu's native attention.
llm.npu supports two types of attentions, i.e., full attention on 4 CPU cores (1 prime core and 3 middle cores) and full attention on GPU.
\sys achieves up to 3.0$\times$ lower latency with \textit{only one middle CPU core.}

\noindent \textbf{\sys can work well on various devices.}
We further show the effectiveness of \sys on other devices.
In Table~\ref{tab:on_8gen2}, we show the inference latency of PhoneLM-0.5B on Redmi K60 Champion Edition.
\sys still shows clear speedup against the baseline (2$\times$/1.25$\times$/1.22$\times$, respectively).
The speedup is lower than on MI14.
This is mainly due to that the Redmi K60 Champion Edition is equipped with V73 series NPU, which is inferior to the V75 NPU on MI14.

% \usepackage{booktabs}
\begin{table}[]
\footnotesize
\begin{tabular}{@{}cccccc@{}}
\toprule
\multicolumn{2}{c}{\textbf{ArxivSum}} & \multicolumn{2}{c}{\textbf{Octopus}} & \multicolumn{2}{c}{\textbf{DroidCall}} \\ \midrule
C/G-Full          & Ours              & C/G-Full          & Ours             & C/G-Full           & Ours              \\ \midrule
1.2 Min.          & 0.6 Min.          & 2.5 Min.          & 2.0 Min.         & 2.7 Min.           & 2.2 Min.          \\ \bottomrule
\end{tabular}
\caption{Running \sys on Redmi K60 Champion Edition (Snapdragon 8Gen2). Model: PhoneLM-0.5B.}
\label{tab:on_8gen2}
\end{table}

\noindent \textbf{\sys is energy efficient.}
We show the energy consumption of running a single attention kernel with an input length of 1024 on Redmi K60 Champion Edition.
The results are obtained from system APIs\footnote{/sys/class/power\_supply/battery/voltage\_now}\footnote{/sys/class/power\_supply/battery/current\_now}.
As shown in Table~\ref{tab:energy}, \sys exhibits up to 7.66$\times$ lower energy consumption.
This is primarily due to the reduced computational load and the proper utilization of NPU, which has relatively low power.

% \usepackage{booktabs}
\begin{table}[]
\footnotesize
\begin{tabular}{@{}cccccc@{}}
\toprule
\textbf{\begin{tabular}[c]{@{}c@{}}Model\\ (Single Kernel)\end{tabular}} &
  \textbf{\begin{tabular}[c]{@{}c@{}}C/G-\\ Full\end{tabular}} &
  \textbf{\begin{tabular}[c]{@{}c@{}}C/G-\\ Sparse\end{tabular}} &
  \textbf{\begin{tabular}[c]{@{}c@{}}C/G-Block-\\ Sparse\end{tabular}} &
  \textbf{\begin{tabular}[c]{@{}c@{}}NPU-\\ Full\end{tabular}} &
  \textbf{Ours} \\ \midrule
PhoneLM-0.5B & 3.72 J & 2.81 J & 1.10 J & 0.85 J & 0.66 J \\ \midrule
PhoneLM-1.5B & 8.59 J & 5.16 J & 1.51 J & 1.29 J & 1.12 J \\ \midrule
Qwen2-0.5B   & 3.38 J & 2.35 J & 0.88 J & 0.85 J & 0.56 J \\ \midrule
Qwen2-1.5B   & 5.29 J & 3.28 J & 0.85 J & 1.10 J & 0.71 J \\ \bottomrule
\end{tabular}
\caption{Energy consumption of an attention kernel. Device: Redmi K60 Champion Edition; Input length: 1024.}
\label{tab:energy}
\end{table}

\subsection{Sensitivity Analysis and Ablation Study}
\label{subsec:sensitivity_ablation}

\begin{figure}[t]
    \centering
      \begin{minipage}[b]{0.23\textwidth}
        \includegraphics[width=\textwidth]{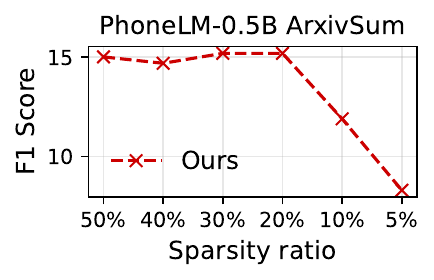}
        \subcaption{Sparsity ratio v.s. accuracy.}
        \label{fig:ratio_acc}
      \end{minipage}
      % \hfill
      \begin{minipage}[b]{0.23\textwidth}
        \includegraphics[width=\textwidth]{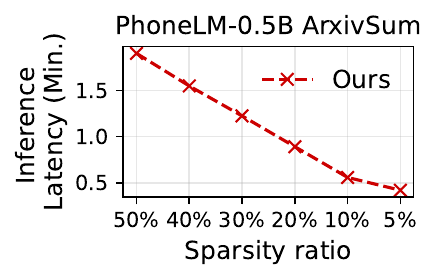}
        \subcaption{Sparsity ratio v.s. latency.}
        \label{fig:ratio_latency}
      \end{minipage}
    \caption{Sensitivity analysis of global sparsity ratio.}
    \label{fig:ratios}

\end{figure}

\begin{figure}[t]
    \centering
      \begin{minipage}[b]{0.23\textwidth}
        \includegraphics[width=\textwidth]{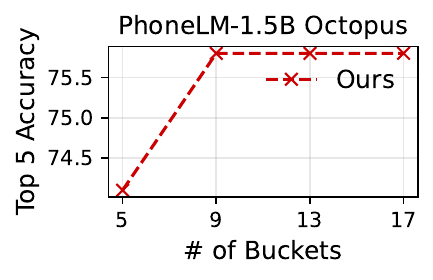}
        \subcaption{Sensitivity to bucket number.}
        \label{fig:bucket_number}
      \end{minipage}
      % \hfill
      \begin{minipage}[b]{0.23\textwidth}
        \includegraphics[width=\textwidth]{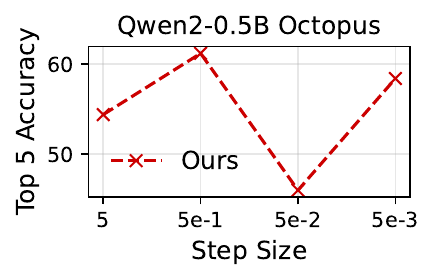}
        \subcaption{Sensitivity to step size.}
        \label{fig:stepsize}
      \end{minipage}
    \caption{Sensitivity analysis of scale factor buckets.}
    \label{fig:buckets}

\end{figure}

\begin{figure}[t]
    \centering
    \includegraphics[width=0.48\textwidth]{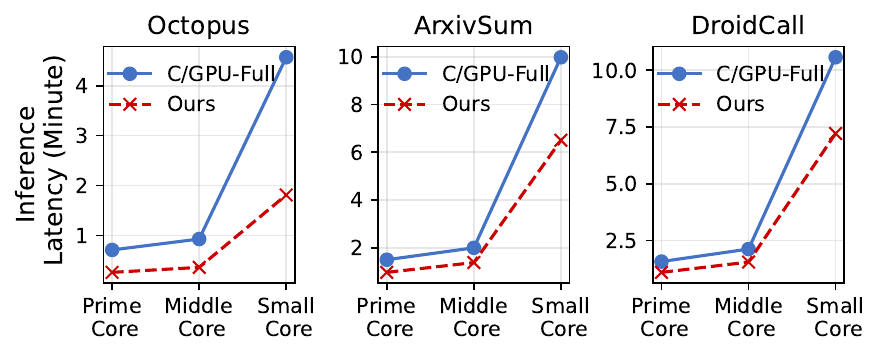}
    \caption{Varying the available resource of CPU/GPU.}
    \label{fig:FSUs}
\end{figure}

\begin{figure}[t]
    \centering
    \includegraphics[width=0.48\textwidth]{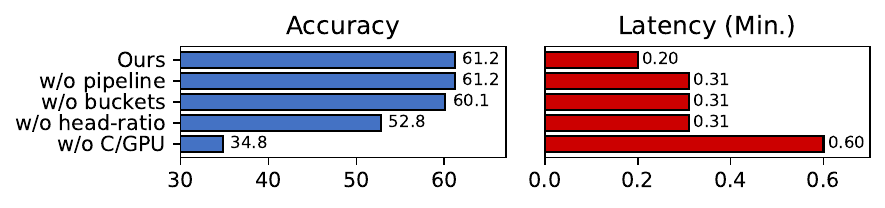}
    \caption{Ablation study on Qwen2-0.5B, MI14, Octopus.}
    \label{fig:ablation}
\end{figure}

\noindent \textbf{Global sparsity ratio.}
We show the sensitivity of sparsity ratio in Figure~\ref{fig:ratios}.
Setting the ratio to 20\% is a reasonable choice.
A too large ratio (e.g. 30\%, 40\%, or 50\%) reduces less latency;
a too small ratio leads to significant accuracy loss.
20\% is the minimum sparsity ratio that maintains no accuracy loss, i.e., the knee point.
As a result, \sys chooses 20\% as its default setting, which also aligns well with the aforementioned analysis in $\S$\ref{sec:design}.

\noindent \textbf{Scale factor buckets.}
Recall that we cache subgraphs of various scales on NPU for Q$\cdot$K.
In \sys's default setting, the number of buckets is set to 9 with a step size of 5e-1.
More buckets offer a more fine-grained tradeoff space, which leads to higher accuracy (5 buckets to 9 buckets in Figure~\ref{fig:bucket_number}); yet the number of buckets exhibits a marginal effect that adding more buckets shows almost no performance gain.
As a result, typically we choose it as 9.
Figure~\ref{fig:stepsize} shows why we choose a step size of 5e-1.
Factors of other orders of magnitude lead to significant performance degradation.
Interestingly, a very tiny step size (5e-3) results in less accuracy drop (yet still significant).
This is primarily because that such a choice makes all samples fall back into the original bucket, since the scale factors of others are too large or too small.
In a nut shell, the default setting of \sys on the buckets is reasonable and leads to the best performance.

\noindent \textbf{The available CPU/GPU.}
Recall that in our default setting, we set the available CPU/GPU resource to a middle core of ARM CPU.
We now vary the available resource in Figure~\ref{fig:FSUs} by mapping it to cores with various compute capability.
The experiment is run on MI14, which has 1 prime core, 5 middle cores, and 2 small cores.
Firstly, \sys consistently outperforms C/G-Full.
Besides, when expanding the compute resource (e.g., small core to middle core to prime core), the absolute inference latency of \sys is also reduced, showing strong scalability.
Overall, \sys is a robust solution under various amount of compute resource.
% To this end, we suggest a strong FSU in the next generation mobile NPUs to maximize the performance. 

\noindent \textbf{Ablation study.}
We conduct ablation study by gradually removing the key designs of \sys.
The results are shown in Figure~\ref{fig:ablation}.
When the pipeline is removed from \sys, the end-to-end latency exhibits a significant rise.
This is because that the pipeline overlaps the latency of NPU and CPU/GPU.
On top of this, when we further remove the design of scale factor buckets, the accuracy shows a clear drop (1.2 pp).
This is because the buckets allow the NPU to adapt better to the input range.
Further, when we make each head share the same global sparsity ratio, the accuracy drops from 60.1 pp to 52.8 pp.
The rationale is that the intrinsic importance of each head is ignored.
Finally, when we even cancel the design of sparse compute on CPU/GPU and directly offloading the full attention to NPU, the accuracy is lowered to 34.8 pp, and the latency is moved to 0.6 minutes.
This is because \sys have to suffer the quantization loss on NPU, and the NPU is not friendly to sparse compute.
In conclusion, each part of \sys shows non-trivial contribution to its performance.

% \subsection{Running alongside Other Mobile Apps}

% \input{fig-with_other_apps}

% \usepackage{booktabs}
\begin{table}[]
\footnotesize
\begin{tabular}{@{}cccccc@{}}
\toprule
\textbf{App}     & \circled{1} & \circled{2} & \circled{3} & \circled{4} & \circled{5} \\ \midrule
\textbf{Latency} & 14.7 Sec.   & 14.5 Sec.   & 15.5 Sec.   & 14.7 Sec.   & 15.2 Sec.   \\ \bottomrule
\end{tabular}
\caption{Running \sys with other apps. Model: PhoneLM-0.5B; Dataset: Octopus; Device: MI14. \protect\circled{1}: No concurrent apps; \protect\circled{2}: Taking photos; \protect\circled{3}: Watching videos; \protect\circled{4}: Listening to music; \protect\circled{5}: Browsing shopping apps.}
\label{tab:with_other_apps}
\end{table}

\noindent \textbf{Running alongside other mobile apps.}
We show \sys's superiority by running the LLM inference together with other mobile apps.
Specifically, we test the end-to-end inference time when the smartphone user is watching videos (GPU and CPU busy), listening to music (CPU busy), tasking photos (ISP and CPU busy) and browsing shopping apps (CPU busy).
We exclude other apps for running on the selected middle core.
The results are shown in Table~\ref{tab:with_other_apps}.
The LLM inference can work well with these apps, with no obvious performance degradation.
% We believe that with \sys, the LLM inference workload will become even more transparent to OS/users and less intrusive to other mobile apps.
\section{Related Work}

\noindent \textbf{On-Device LLM Inference.}
Running mobile LLMs~\cite{zhang2024tinyllamaopensourcesmalllanguage, thawakar2024mobillamaaccuratelightweightfully, ma2024era1bitllmslarge, zhang2022optopenpretrainedtransformer, abdin2024phi3technicalreporthighly, lee2024exploreselectderiverecall, subramanian2025smalllanguagemodelsslms, 10.1145/3662006.3662059} on devices has become a key area in privacy-preserving mobile intelligence~\cite{navardi2025genaiedgecomprehensivesurvey, belcak2025smalllanguagemodelsfuture, li2025mobillmenablingllmfinetuning, skliar2025mixturecacheconditionalexpertsefficient, peng2024pocketllmenablingondevicefinetuning}.
There exist many popular frameworks.
For instance, HeteroLLM~\cite{chen2025heterollmacceleratinglargelanguage} proposes a GPU/NPU collaborated framework for LLM inference on mobile SoCs.
It optimizes the memory copy efficiency.
However, the attention and parts of MLP still run on GPU, which can be busy during the usage of mobile devices.
Akin to HeteroLLM, llm.npu~\cite{llm.npu} runs attention on CPU/GPU, which cannot realize the NPU-centric inference.
PowerinferV2~\cite{xue2024powerinfer2fastlargelanguage} employs dense attention on NPUs, i.e., the NPU-Full baseline.
However, compared to \sys, it does not support sparse attention.
Even worse, it still suffers from the accuracy drop of attention quantization, especially on mobile-sized LLMs and mobile agent tasks.
Its main focus is on extending the RAM of mobile devices to run large-scale LLMs that have stronger resilience to attention quantization.
Akin to PowerinferV2, QNN~\cite{qnn} also runs full attention on NPU, encountering the same problems.
There are also frameworks that run both the attention and non-attention parts on CPU/GPU.
For example, MLC-LLM~\cite{mlc-llm} and TFLite~\cite{mediapipe} run LLMs on GPUs; Llama.cpp~\cite{llamacpp} runs LLMs on CPUs.
By only keeping the sparse compute of important tokens on CPU/GPU, \sys realizes the most accurate and efficient NPU-centric inference.

\noindent \textbf{Efficient Sparse Attention.}
Fixed-pattern sparse attention~\cite{jiang2024minference10acceleratingprefilling, jiang2023mistral7b, lai2025flexprefillcontextawaresparseattention, xiao2024duoattentionefficientlongcontextllm} identifies one of given patterns offline.
Common patterns include sliding window, $\Lambda$-shape and vertical-slash, etc.
Compared to ad-hoc fixed patterns, \sys's dynamic attention is a superset of the above patterns.
Block sparse attention~\cite{yang2025lserveefficientlongsequencellm, zhang2025spargeattn, tang2024questqueryawaresparsityefficient, xu2025xattentionblocksparseattention} pre-pools the Q and K, and selects top k areas in coarse-grained block level.
As discussed before, such a paradigm is less accurate.
Reserving/discarding a whole region may dismiss important tokens.
\sys's fine-grained sparse attention is much more accurate.
Recently there is also a trend of training native sparse attentions~\cite{yuan2025nativesparseattentionhardwarealigned, lu2025mobamixtureblockattention}.
Such methods make the attention inherently sparse at the inference stage.
However, the learned sparsity cannot directly transfer to downstream tasks and fine-tuned models.
The resource consumption of pre-training is also not affordable to each developer.
In comparison, the sparse attention of \sys is plug-and-play.

\noindent \textbf{Mobile Intelligence.}
Mobile intelligence runs DNNs especially foundation models on edge/mobile devices for sensing the human/environment or acting with the cyber/physical world~\cite{10.1145/3706418, 10812936, yin2024llmservicemobiledevices, yin2024elmselasticizedlargelanguage, baris2025foundationmodelscpsiotopportunities, xu2024surveyresourceefficientllmmultimodal, ouyang2024admarkermultimodalfederatedlearning, TaskSense, yang2025contextagentcontextawareproactivellm, 10.1145/3666025.3699327}.
With \sys's NPU-centric inference, the execution of foundation models will be more transparent and less intrusive to the human and environment.

\section{Conclusion}

We present \sys, a sparse attention module for NPU-centric on-device LLM inference.
It incorporates innovative techniques including NPU-based estimation, NPU compute graph bucketing and NPU-CPU/GPU pipeline.
It losslessly achieves up to 4.5$\times$ end-to-end speed up while minimizing the reliance on CPU/GPU resources.

% use the ACM bibliography style
\bibliographystyle{ACM-Reference-Format}
\bibliography{ref-yws}

@misc{yang2025lserveefficientlongsequencellm,
      title={LServe: Efficient Long-sequence LLM Serving with Unified Sparse Attention}, 
      author={Shang Yang and Junxian Guo and Haotian Tang and Qinghao Hu and Guangxuan Xiao and Jiaming Tang and Yujun Lin and Zhijian Liu and Yao Lu and Song Han},
      year={2025},
      eprint={2502.14866},
      archivePrefix={arXiv},
      primaryClass={cs.CL},
      url={https://arxiv.org/abs/2502.14866}, 
}

@inproceedings{llm.npu,
author = {Xu, Daliang and Zhang, Hao and Yang, Liming and Liu, Ruiqi and Huang, Gang and Xu, Mengwei and Liu, Xuanzhe},
title = {Fast On-device LLM Inference with NPUs},
year = {2025},
isbn = {9798400706981},
publisher = {Association for Computing Machinery},
address = {New York, NY, USA},
url = {https://doi.org/10.1145/3669940.3707239},
doi = {10.1145/3669940.3707239},
booktitle = {Proceedings of the 30th ACM International Conference on Architectural Support for Programming Languages and Operating Systems, Volume 1},
pages = {445–462},
numpages = {18},
location = {Rotterdam, Netherlands},
series = {ASPLOS '25}
}

@misc{chen2025heterollmacceleratinglargelanguage,
      title={HeteroLLM: Accelerating Large Language Model Inference on Mobile SoCs platform with Heterogeneous AI Accelerators}, 
      author={Le Chen and Dahu Feng and Erhu Feng and Rong Zhao and Yingrui Wang and Yubin Xia and Haibo Chen and Pinjie Xu},
      year={2025},
      eprint={2501.14794},
      archivePrefix={arXiv},
      primaryClass={cs.DC},
      url={https://arxiv.org/abs/2501.14794}, 
}

@misc{xue2024powerinfer2fastlargelanguage,
      title={PowerInfer-2: Fast Large Language Model Inference on a Smartphone}, 
      author={Zhenliang Xue and Yixin Song and Zeyu Mi and Xinrui Zheng and Yubin Xia and Haibo Chen},
      year={2024},
      eprint={2406.06282},
      archivePrefix={arXiv},
      primaryClass={cs.LG},
      url={https://arxiv.org/abs/2406.06282}, 
}

@Misc{qnn,
	title           = {QNN SDK.},
	howpublished    = {\url{https://docs.qualcomm.com/bundle/publicresource/topics/80-63442-50/introduction.html}},
        year = 2025
}

@software{mlc-llm,
    author = {{MLC team}},
    title = {{MLC-LLM}},
    url = {https://github.com/mlc-ai/mlc-llm},
    year = {2023-2025}
}

@software{mediapipe,
    author = {{TFLite team}},
    title = {{mediapipe}},
    url = {https://ai.google.dev/edge/mediapipe/solutions/guide},
    year = {2025}
}

@software{llamacpp,
    author = {{ggml}},
    title = {{llama.cpp}},
    url = {https://github.com/ggml-org/llama.cpp},
    year = {2025}
}

@misc{lai2025flexprefillcontextawaresparseattention,
      title={FlexPrefill: A Context-Aware Sparse Attention Mechanism for Efficient Long-Sequence Inference}, 
      author={Xunhao Lai and Jianqiao Lu and Yao Luo and Yiyuan Ma and Xun Zhou},
      year={2025},
      eprint={2502.20766},
      archivePrefix={arXiv},
      primaryClass={cs.LG},
      url={https://arxiv.org/abs/2502.20766}, 
}

@misc{jiang2023mistral7b,
      title={Mistral 7B}, 
      author={Albert Q. Jiang and Alexandre Sablayrolles and Arthur Mensch and Chris Bamford and Devendra Singh Chaplot and Diego de las Casas and Florian Bressand and Gianna Lengyel and Guillaume Lample and Lucile Saulnier and Lélio Renard Lavaud and Marie-Anne Lachaux and Pierre Stock and Teven Le Scao and Thibaut Lavril and Thomas Wang and Timothée Lacroix and William El Sayed},
      year={2023},
      eprint={2310.06825},
      archivePrefix={arXiv},
      primaryClass={cs.CL},
      url={https://arxiv.org/abs/2310.06825}, 
}

@misc{jiang2024minference10acceleratingprefilling,
      title={MInference 1.0: Accelerating Pre-filling for Long-Context LLMs via Dynamic Sparse Attention}, 
      author={Huiqiang Jiang and Yucheng Li and Chengruidong Zhang and Qianhui Wu and Xufang Luo and Surin Ahn and Zhenhua Han and Amir H. Abdi and Dongsheng Li and Chin-Yew Lin and Yuqing Yang and Lili Qiu},
      year={2024},
      eprint={2407.02490},
      archivePrefix={arXiv},
      primaryClass={cs.CL},
      url={https://arxiv.org/abs/2407.02490}, 
}

@misc{yuan2025nativesparseattentionhardwarealigned,
      title={Native Sparse Attention: Hardware-Aligned and Natively Trainable Sparse Attention}, 
      author={Jingyang Yuan and Huazuo Gao and Damai Dai and Junyu Luo and Liang Zhao and Zhengyan Zhang and Zhenda Xie and Y. X. Wei and Lean Wang and Zhiping Xiao and Yuqing Wang and Chong Ruan and Ming Zhang and Wenfeng Liang and Wangding Zeng},
      year={2025},
      eprint={2502.11089},
      archivePrefix={arXiv},
      primaryClass={cs.CL},
      url={https://arxiv.org/abs/2502.11089}, 
}

@article{10.1145/3706418,
author = {Xu, Mengwei and Cai, Dongqi and Yin, Wangsong and Wang, Shangguang and Jin, Xin and Liu, Xuanzhe},
title = {Resource-efficient Algorithms and Systems of Foundation Models: A Survey},
year = {2025},
issue_date = {May 2025},
publisher = {Association for Computing Machinery},
address = {New York, NY, USA},
volume = {57},
number = {5},
issn = {0360-0300},
url = {https://doi.org/10.1145/3706418},
doi = {10.1145/3706418},
journal = {ACM Comput. Surv.},
month = jan,
articleno = {110},
numpages = {39}
}

@ARTICLE{10812936,
  author={Xu, Daliang and Yin, Wangsong and Zhang, Hao and Jin, Xin and Zhang, Ying and Wei, Shiyun and Xu, Mengwei and Liu, Xuanzhe},
  journal={IEEE Transactions on Mobile Computing}, 
  title={EdgeLLM: Fast On-Device LLM Inference With Speculative Decoding}, 
  year={2025},
  volume={24},
  number={4},
  pages={3256-3273},
  doi={10.1109/TMC.2024.3513457}}

@misc{yin2024llmservicemobiledevices,
      title={LLM as a System Service on Mobile Devices}, 
      author={Wangsong Yin and Mengwei Xu and Yuanchun Li and Xuanzhe Liu},
      year={2024},
      eprint={2403.11805},
      archivePrefix={arXiv},
      primaryClass={cs.OS},
      url={https://arxiv.org/abs/2403.11805}, 
}

@misc{yin2024elmselasticizedlargelanguage,
      title={ELMS: Elasticized Large Language Models On Mobile Devices}, 
      author={Wangsong Yin and Rongjie Yi and Daliang Xu and Gang Huang and Mengwei Xu and Xuanzhe Liu},
      year={2024},
      eprint={2409.09071},
      archivePrefix={arXiv},
      primaryClass={cs.DC},
      url={https://arxiv.org/abs/2409.09071}, 
}

@misc{baris2025foundationmodelscpsiotopportunities,
      title={Foundation Models for CPS-IoT: Opportunities and Challenges}, 
      author={Ozan Baris and Yizhuo Chen and Gaofeng Dong and Liying Han and Tomoyoshi Kimura and Pengrui Quan and Ruijie Wang and Tianchen Wang and Tarek Abdelzaher and Mario Bergés and Paul Pu Liang and Mani Srivastava},
      year={2025},
      eprint={2501.16368},
      archivePrefix={arXiv},
      primaryClass={cs.LG},
      url={https://arxiv.org/abs/2501.16368}, 
}

@article{Mersha_2024,
   title={Explainable artificial intelligence: A survey of needs, techniques, applications, and future direction},
   volume={599},
   ISSN={0925-2312},
   url={http://dx.doi.org/10.1016/j.neucom.2024.128111},
   DOI={10.1016/j.neucom.2024.128111},
   journal={Neurocomputing},
   publisher={Elsevier BV},
   author={Mersha, Melkamu and Lam, Khang and Wood, Joseph and AlShami, Ali K. and Kalita, Jugal},
   year={2024},
   month=sep, pages={128111} }

@article{Cai_2025,
   title={A Survey on Mixture of Experts in Large Language Models},
   ISSN={2326-3865},
   url={http://dx.doi.org/10.1109/TKDE.2025.3554028},
   DOI={10.1109/tkde.2025.3554028},
   journal={IEEE Transactions on Knowledge and Data Engineering},
   publisher={Institute of Electrical and Electronics Engineers (IEEE)},
   author={Cai, Weilin and Jiang, Juyong and Wang, Fan and Tang, Jing and Kim, Sunghun and Huang, Jiayi},
   year={2025},
   pages={1–20} }

@misc{ge2024modeltellsdiscardadaptive,
      title={Model Tells You What to Discard: Adaptive KV Cache Compression for LLMs}, 
      author={Suyu Ge and Yunan Zhang and Liyuan Liu and Minjia Zhang and Jiawei Han and Jianfeng Gao},
      year={2024},
      eprint={2310.01801},
      archivePrefix={arXiv},
      primaryClass={cs.CL},
      url={https://arxiv.org/abs/2310.01801}, 
}

@software{mi14,
    author = {{xiaomi}},
    title = {{MI14 Smartphone}},
    url = {https://www.mi.com/global/product/xiaomi-14/specs/},
    year = {2025}
}

@software{k60,
    author = {{redmi}},
    title = {{Redmi K60 Champion Edition Smartphone}},
    url = {https://www.gsmarena.com/xiaomi_redmi_k60_pro-12046.php},
    year = {2025}
}

@software{qwen2-0.5b,
    author = {{qwen}},
    title = {{Qwen2-0.5B}},
    url = {https://huggingface.co/unsloth/Qwen2-0.5B},
    year = {2025}
}

@software{qwen2-1.5b,
    author = {{qwen}},
    title = {{Qwen2-1.5B}},
    url = {https://huggingface.co/unsloth/Qwen2-1.5B},
    year = {2025}
}

@software{phonelm-0.5b,
    author = {{phonelm}},
    title = {{PhoneLM-0.5B}},
    url = {https://huggingface.co/unsloth/PhoneLM-0.5B},
    year = {2025}
}

@software{phonelm-1.5b,
    author = {{phonelm}},
    title = {{PhoneLM-1.5B}},
    url = {https://huggingface.co/unsloth/PhoneLM-1.5B},
    year = {2025}
}

@inproceedings{cohan-etal-2018-discourse,
  title = "A Discourse-Aware Attention Model for Abstractive Summarization of Long Documents",
  author = "Cohan, Arman  and
    Dernoncourt, Franck  and
    Kim, Doo Soon  and
    Bui, Trung  and
    Kim, Seokhwan  and
    Chang, Walter  and
    Goharian, Nazli",
  booktitle = "Proceedings of the 2018 Conference of the North {A}merican Chapter of the Association for Computational Linguistics: Human Language Technologies, Volume 2 (Short Papers)",
  month = jun,
  year = "2018",
  address = "New Orleans, Louisiana",
  publisher = "Association for Computational Linguistics",
  url = "https://aclanthology.org/N18-2097",
  doi = "10.18653/v1/N18-2097",
  pages = "615--621",
}

@misc{xie2024droidcalldatasetllmpoweredandroid,
      title={DroidCall: A Dataset for LLM-powered Android Intent Invocation}, 
      author={Weikai Xie and Li Zhang and Shihe Wang and Rongjie Yi and Mengwei Xu},
      year={2024},
      eprint={2412.00402},
      archivePrefix={arXiv},
      primaryClass={cs.AI},
      url={https://arxiv.org/abs/2412.00402}, 
}

@misc{chen2024octopusv2ondevicelanguage,
      title={Octopus v2: On-device language model for super agent}, 
      author={Wei Chen and Zhiyuan Li},
      year={2024},
      eprint={2404.01744},
      archivePrefix={arXiv},
      primaryClass={cs.CL},
      url={https://arxiv.org/abs/2404.01744}, 
}

@article{wang2021spatten,
        title={SpAtten: Efficient Sparse Attention Architecture with Cascade Token and Head Pruning},
        author={Wang, Hanrui and Zhang, Zhekai and Han, Song},
        journal={HPCA},
        year={2021}
        }

@misc{zhang2023h2oheavyhitteroracleefficient,
      title={H$_2$O: Heavy-Hitter Oracle for Efficient Generative Inference of Large Language Models}, 
      author={Zhenyu Zhang and Ying Sheng and Tianyi Zhou and Tianlong Chen and Lianmin Zheng and Ruisi Cai and Zhao Song and Yuandong Tian and Christopher Ré and Clark Barrett and Zhangyang Wang and Beidi Chen},
      year={2023},
      eprint={2306.14048},
      archivePrefix={arXiv},
      primaryClass={cs.LG},
      url={https://arxiv.org/abs/2306.14048}, 
}

@misc{bai2023qwentechnicalreport,
      title={Qwen Technical Report}, 
      author={Jinze Bai and Shuai Bai and Yunfei Chu and Zeyu Cui and Kai Dang and Xiaodong Deng and Yang Fan and Wenbin Ge and Yu Han and Fei Huang and Binyuan Hui and Luo Ji and Mei Li and Junyang Lin and Runji Lin and Dayiheng Liu and Gao Liu and Chengqiang Lu and Keming Lu and Jianxin Ma and Rui Men and Xingzhang Ren and Xuancheng Ren and Chuanqi Tan and Sinan Tan and Jianhong Tu and Peng Wang and Shijie Wang and Wei Wang and Shengguang Wu and Benfeng Xu and Jin Xu and An Yang and Hao Yang and Jian Yang and Shusheng Yang and Yang Yao and Bowen Yu and Hongyi Yuan and Zheng Yuan and Jianwei Zhang and Xingxuan Zhang and Yichang Zhang and Zhenru Zhang and Chang Zhou and Jingren Zhou and Xiaohuan Zhou and Tianhang Zhu},
      year={2023},
      eprint={2309.16609},
      archivePrefix={arXiv},
      primaryClass={cs.CL},
      url={https://arxiv.org/abs/2309.16609}, 
}

@misc{qwen2025qwen25technicalreport,
      title={Qwen2.5 Technical Report}, 
      author={Qwen and : and An Yang and Baosong Yang and Beichen Zhang and Binyuan Hui and Bo Zheng and Bowen Yu and Chengyuan Li and Dayiheng Liu and Fei Huang and Haoran Wei and Huan Lin and Jian Yang and Jianhong Tu and Jianwei Zhang and Jianxin Yang and Jiaxi Yang and Jingren Zhou and Junyang Lin and Kai Dang and Keming Lu and Keqin Bao and Kexin Yang and Le Yu and Mei Li and Mingfeng Xue and Pei Zhang and Qin Zhu and Rui Men and Runji Lin and Tianhao Li and Tianyi Tang and Tingyu Xia and Xingzhang Ren and Xuancheng Ren and Yang Fan and Yang Su and Yichang Zhang and Yu Wan and Yuqiong Liu and Zeyu Cui and Zhenru Zhang and Zihan Qiu},
      year={2025},
      eprint={2412.15115},
      archivePrefix={arXiv},
      primaryClass={cs.CL},
      url={https://arxiv.org/abs/2412.15115}, 
}

@Misc{rewind,
	title           = {rewind},
	howpublished    = {\url{https://www.rewind.ai/}},
	year            = {2025}
}

@misc{zhang2024llamatouch,
      title={LlamaTouch: A Faithful and Scalable Testbed for Mobile UI Task Automation}, 
      author={Li Zhang and Shihe Wang and Xianqing Jia and Zhihan Zheng and Yunhe Yan and Longxi Gao and Yuanchun Li and Mengwei Xu},
      year={2024},
      eprint={2404.16054},
      archivePrefix={arXiv},
      primaryClass={cs.HC},
      url={https://arxiv.org/abs/2404.16054}, 
}

@Misc{8gen3,
	title           = {Snapdragon 8 gen 3 mobile platform product brief.},
	howpublished    = {\url{https://docs.qualcomm.com/bundle/publicresource/87-71408-1_REV_C_Snapdragon_8_gen_3_Mobile_Platform_Product_Brief.pdf}},
        year = 2025
}

@Misc{TMS320F2812,
	title           = {TMS320F2812 platform product brief.},
	howpublished    = {\url{https://www.ti.com/product/TMS320F2812}},
        year = 2025
}

@Misc{orin,
	title           = {Nvidia Jetson Orin},
	howpublished    = {\url{https://www.nvidia.com/en-us/autonomous-machines/embedded-systems/jetson-orin/}},
        year = 2025
}

@misc{xiao2024smoothquantaccurateefficientposttraining,
      title={SmoothQuant: Accurate and Efficient Post-Training Quantization for Large Language Models}, 
      author={Guangxuan Xiao and Ji Lin and Mickael Seznec and Hao Wu and Julien Demouth and Song Han},
      year={2024},
      eprint={2211.10438},
      archivePrefix={arXiv},
      primaryClass={cs.CL},
      url={https://arxiv.org/abs/2211.10438}, 
}

@misc{lin2024awqactivationawareweightquantization,
      title={AWQ: Activation-aware Weight Quantization for LLM Compression and Acceleration}, 
      author={Ji Lin and Jiaming Tang and Haotian Tang and Shang Yang and Wei-Ming Chen and Wei-Chen Wang and Guangxuan Xiao and Xingyu Dang and Chuang Gan and Song Han},
      year={2024},
      eprint={2306.00978},
      archivePrefix={arXiv},
      primaryClass={cs.CL},
      url={https://arxiv.org/abs/2306.00978}, 
}

@ARTICLE{9916240,
  author={Park, Jun-Seok and Park, Changsoo and Kwon, Suknam and Jeon, Taeho and Kang, Yesung and Lee, Heonsoo and Lee, Dongwoo and Kim, James and Kim, Hyeong-Seok and Lee, YoungJong and Park, Sangkyu and Kim, MinSeong and Ha, SangHyuck and Bang, Jihoon and Park, Jinpyo and Lim, SukHwan and Kang, Inyup},
  journal={IEEE Journal of Solid-State Circuits}, 
  title={A Multi-Mode 8k-MAC HW-Utilization-Aware Neural Processing Unit With a Unified Multi-Precision Datapath in 4-nm Flagship Mobile SoC}, 
  year={2023},
  volume={58},
  number={1},
  pages={189-202},
  doi={10.1109/JSSC.2022.3205713}}

@misc{lokhande2025polaronprecisionawareondevicelearning,
      title={POLARON: Precision-aware On-device Learning and Adaptive Runtime-cONfigurable AI acceleration}, 
      author={Mukul Lokhande and Santosh Kumar Vishvakarma},
      year={2025},
      eprint={2506.08785},
      archivePrefix={arXiv},
      primaryClass={cs.AR},
      url={https://arxiv.org/abs/2506.08785}, 
}

@misc{lokhande2024flexpeflexiblesimdmultiprecision,
      title={Flex-PE: Flexible and SIMD Multi-Precision Processing Element for AI Workloads}, 
      author={Mukul Lokhande and Gopal Raut and Santosh Kumar Vishvakarma},
      year={2024},
      eprint={2412.11702},
      archivePrefix={arXiv},
      primaryClass={cs.AR},
      url={https://arxiv.org/abs/2412.11702}, 
}

@misc{merity2016pointer,
      title={Pointer Sentinel Mixture Models},
      author={Stephen Merity and Caiming Xiong and James Bradbury and Richard Socher},
      year={2016},
      eprint={1609.07843},
      archivePrefix={arXiv},
      primaryClass={cs.CL}
}

@misc{xu2024surveyresourceefficientllmmultimodal,
      title={A Survey of Resource-efficient LLM and Multimodal Foundation Models}, 
      author={Mengwei Xu and Wangsong Yin and Dongqi Cai and Rongjie Yi and Daliang Xu and Qipeng Wang and Bingyang Wu and Yihao Zhao and Chen Yang and Shihe Wang and Qiyang Zhang and Zhenyan Lu and Li Zhang and Shangguang Wang and Yuanchun Li and Yunxin Liu and Xin Jin and Xuanzhe Liu},
      year={2024},
      eprint={2401.08092},
      archivePrefix={arXiv},
      primaryClass={cs.LG},
      url={https://arxiv.org/abs/2401.08092}, 
}

@misc{xiao2024duoattentionefficientlongcontextllm,
      title={DuoAttention: Efficient Long-Context LLM Inference with Retrieval and Streaming Heads}, 
      author={Guangxuan Xiao and Jiaming Tang and Jingwei Zuo and Junxian Guo and Shang Yang and Haotian Tang and Yao Fu and Song Han},
      year={2024},
      eprint={2410.10819},
      archivePrefix={arXiv},
      primaryClass={cs.CL},
      url={https://arxiv.org/abs/2410.10819}, 
}

@inproceedings{zhang2025spargeattn, title={Spargeattn: Accurate sparse attention accelerating any model inference}, author={Zhang, Jintao and Xiang, Chendong and Huang, Haofeng and Wei, Jia and Xi, Haocheng and Zhu, Jun and Chen, Jianfei}, booktitle={International Conference on Machine Learning (ICML)}, year={2025} }

@misc{tang2024questqueryawaresparsityefficient,
      title={Quest: Query-Aware Sparsity for Efficient Long-Context LLM Inference}, 
      author={Jiaming Tang and Yilong Zhao and Kan Zhu and Guangxuan Xiao and Baris Kasikci and Song Han},
      year={2024},
      eprint={2406.10774},
      archivePrefix={arXiv},
      primaryClass={cs.CL},
      url={https://arxiv.org/abs/2406.10774}, 
}

@misc{xu2025xattentionblocksparseattention,
      title={XAttention: Block Sparse Attention with Antidiagonal Scoring}, 
      author={Ruyi Xu and Guangxuan Xiao and Haofeng Huang and Junxian Guo and Song Han},
      year={2025},
      eprint={2503.16428},
      archivePrefix={arXiv},
      primaryClass={cs.CL},
      url={https://arxiv.org/abs/2503.16428}, 
}

@misc{lu2025mobamixtureblockattention,
      title={MoBA: Mixture of Block Attention for Long-Context LLMs}, 
      author={Enzhe Lu and Zhejun Jiang and Jingyuan Liu and Yulun Du and Tao Jiang and Chao Hong and Shaowei Liu and Weiran He and Enming Yuan and Yuzhi Wang and Zhiqi Huang and Huan Yuan and Suting Xu and Xinran Xu and Guokun Lai and Yanru Chen and Huabin Zheng and Junjie Yan and Jianlin Su and Yuxin Wu and Neo Y. Zhang and Zhilin Yang and Xinyu Zhou and Mingxing Zhang and Jiezhong Qiu},
      year={2025},
      eprint={2502.13189},
      archivePrefix={arXiv},
      primaryClass={cs.LG},
      url={https://arxiv.org/abs/2502.13189}, 
}

@Misc{hexagon_sdk,
	title           = {Hexagon NPU SDK.},
	howpublished    = {\url{https://www.qualcomm.com/developer/software/hexagon-npu-sdk}},
        year = 2025
}

@Misc{arm_neon,
	title           = {ARM NEON.},
	howpublished    = {\url{https://www.arm.com/technologies/neon}},
        year = 2025
}

@Misc{opencl,
	title           = {Open CL.},
	howpublished    = {\url{https://en.wikipedia.org/wiki/OpenCL}},
        year = 2025
}

@Misc{llvm,
	title           = {LLVM.},
	howpublished    = {\url{https://llvm.org/}},
        year = 2025
}

@misc{su2023roformerenhancedtransformerrotary,
      title={RoFormer: Enhanced Transformer with Rotary Position Embedding}, 
      author={Jianlin Su and Yu Lu and Shengfeng Pan and Ahmed Murtadha and Bo Wen and Yunfeng Liu},
      year={2023},
      eprint={2104.09864},
      archivePrefix={arXiv},
      primaryClass={cs.CL},
      url={https://arxiv.org/abs/2104.09864}, 
}

@misc{zhang2024tinyllamaopensourcesmalllanguage,
      title={TinyLlama: An Open-Source Small Language Model}, 
      author={Peiyuan Zhang and Guangtao Zeng and Tianduo Wang and Wei Lu},
      year={2024},
      eprint={2401.02385},
      archivePrefix={arXiv},
      primaryClass={cs.CL},
      url={https://arxiv.org/abs/2401.02385}, 
}

@misc{thawakar2024mobillamaaccuratelightweightfully,
      title={MobiLlama: Towards Accurate and Lightweight Fully Transparent GPT}, 
      author={Omkar Thawakar and Ashmal Vayani and Salman Khan and Hisham Cholakal and Rao M. Anwer and Michael Felsberg and Tim Baldwin and Eric P. Xing and Fahad Shahbaz Khan},
      year={2024},
      eprint={2402.16840},
      archivePrefix={arXiv},
      primaryClass={cs.CL},
      url={https://arxiv.org/abs/2402.16840}, 
}

@misc{ma2024era1bitllmslarge,
      title={The Era of 1-bit LLMs: All Large Language Models are in 1.58 Bits}, 
      author={Shuming Ma and Hongyu Wang and Lingxiao Ma and Lei Wang and Wenhui Wang and Shaohan Huang and Li Dong and Ruiping Wang and Jilong Xue and Furu Wei},
      year={2024},
      eprint={2402.17764},
      archivePrefix={arXiv},
      primaryClass={cs.CL},
      url={https://arxiv.org/abs/2402.17764}, 
}

@misc{zhang2022optopenpretrainedtransformer,
      title={OPT: Open Pre-trained Transformer Language Models}, 
      author={Susan Zhang and Stephen Roller and Naman Goyal and Mikel Artetxe and Moya Chen and Shuohui Chen and Christopher Dewan and Mona Diab and Xian Li and Xi Victoria Lin and Todor Mihaylov and Myle Ott and Sam Shleifer and Kurt Shuster and Daniel Simig and Punit Singh Koura and Anjali Sridhar and Tianlu Wang and Luke Zettlemoyer},
      year={2022},
      eprint={2205.01068},
      archivePrefix={arXiv},
      primaryClass={cs.CL},
      url={https://arxiv.org/abs/2205.01068}, 
}

@misc{abdin2024phi3technicalreporthighly,
      title={Phi-3 Technical Report: A Highly Capable Language Model Locally on Your Phone}, 
      author={Marah Abdin and Jyoti Aneja, etc.},
      year={2024},
      eprint={2404.14219},
      archivePrefix={arXiv},
      primaryClass={cs.CL},
      url={https://arxiv.org/abs/2404.14219}, 
}

@misc{navardi2025genaiedgecomprehensivesurvey,
      title={GenAI at the Edge: Comprehensive Survey on Empowering Edge Devices}, 
      author={Mozhgan Navardi and Romina Aalishah and Yuzhe Fu and Yueqian Lin and Hai Li and Yiran Chen and Tinoosh Mohsenin},
      year={2025},
      eprint={2502.15816},
      archivePrefix={arXiv},
      primaryClass={cs.DC},
      url={https://arxiv.org/abs/2502.15816}, 
}

@misc{belcak2025smalllanguagemodelsfuture,
      title={Small Language Models are the Future of Agentic AI}, 
      author={Peter Belcak and Greg Heinrich and Shizhe Diao and Yonggan Fu and Xin Dong and Saurav Muralidharan and Yingyan Celine Lin and Pavlo Molchanov},
      year={2025},
      eprint={2506.02153},
      archivePrefix={arXiv},
      primaryClass={cs.AI},
      url={https://arxiv.org/abs/2506.02153}, 
}

@misc{li2025mobillmenablingllmfinetuning,
      title={MobiLLM: Enabling LLM Fine-Tuning on the Mobile Device via Server Assisted Side Tuning}, 
      author={Liang Li and Xingke Yang and Wen Wu and Hao Wang and Tomoaki Ohtsuki and Xin Fu and Miao Pan and Xuemin Shen},
      year={2025},
      eprint={2502.20421},
      archivePrefix={arXiv},
      primaryClass={cs.LG},
      url={https://arxiv.org/abs/2502.20421}, 
}

@misc{skliar2025mixturecacheconditionalexpertsefficient,
      title={Mixture of Cache-Conditional Experts for Efficient Mobile Device Inference}, 
      author={Andrii Skliar and Ties van Rozendaal and Romain Lepert and Todor Boinovski and Mart van Baalen and Markus Nagel and Paul Whatmough and Babak Ehteshami Bejnordi},
      year={2025},
      eprint={2412.00099},
      archivePrefix={arXiv},
      primaryClass={cs.LG},
      url={https://arxiv.org/abs/2412.00099}, 
}

@misc{peng2024pocketllmenablingondevicefinetuning,
      title={PocketLLM: Enabling On-Device Fine-Tuning for Personalized LLMs}, 
      author={Dan Peng and Zhihui Fu and Jun Wang},
      year={2024},
      eprint={2407.01031},
      archivePrefix={arXiv},
      primaryClass={cs.LG},
      url={https://arxiv.org/abs/2407.01031}, 
}

@misc{ouyang2024admarkermultimodalfederatedlearning,
      title={ADMarker: A Multi-Modal Federated Learning System for Monitoring Digital Biomarkers of Alzheimer's Disease}, 
      author={Xiaomin Ouyang and Xian Shuai and Yang Li and Li Pan and Xifan Zhang and Heming Fu and Sitong Cheng and Xinyan Wang and Shihua Cao and Jiang Xin and Hazel Mok and Zhenyu Yan and Doris Sau Fung Yu and Timothy Kwok and Guoliang Xing},
      year={2024},
      eprint={2310.15301},
      archivePrefix={arXiv},
      primaryClass={cs.LG},
      url={https://arxiv.org/abs/2310.15301}, 
}

@inbook{TaskSense,
author = {Liu, Kaiwei and Yang, Bufang and Xu, Lilin and Guo, Yunqi and Xing, Guoliang and Shuai, Xian and Ren, Xiaozhe and Jiang, Xin and Yan, Zhenyu},
title = {TaskSense: A Translation-like Approach for Tasking Heterogeneous Sensor Systems with LLMs},
year = {2025},
isbn = {9798400714795},
publisher = {Association for Computing Machinery},
address = {New York, NY, USA},
url = {https://doi.org/10.1145/3715014.3722070},
booktitle = {Proceedings of the 23rd ACM Conference on Embedded Networked Sensor Systems},
pages = {213–225},
numpages = {13}
}

@misc{yang2025contextagentcontextawareproactivellm,
      title={ContextAgent: Context-Aware Proactive LLM Agents with Open-World Sensory Perceptions}, 
      author={Bufang Yang and Lilin Xu and Liekang Zeng and Kaiwei Liu and Siyang Jiang and Wenrui Lu and Hongkai Chen and Xiaofan Jiang and Guoliang Xing and Zhenyu Yan},
      year={2025},
      eprint={2505.14668},
      archivePrefix={arXiv},
      primaryClass={cs.AI},
      url={https://arxiv.org/abs/2505.14668}, 
}

@inproceedings{10.1145/3666025.3699327,
author = {Yin, Wangsong and Xu, Daliang and Huang, Gang and Zhang, Ying and Wei, Shiyun and Xu, Mengwei and Liu, Xuanzhe},
title = {PieBridge: Fast and Parameter-Efficient On-Device Training via Proxy Networks},
year = {2024},
isbn = {9798400706974},
publisher = {Association for Computing Machinery},
address = {New York, NY, USA},
url = {https://doi.org/10.1145/3666025.3699327},
doi = {10.1145/3666025.3699327},
booktitle = {Proceedings of the 22nd ACM Conference on Embedded Networked Sensor Systems},
pages = {126–140},
numpages = {15},
location = {Hangzhou, China},
series = {SenSys '24}
}

@inproceedings{10.1145/3689031.3696067,
author = {Sen, Tanmoy and Shen, Haiying and Iyer, Anand Padmanabha},
title = {Flex: Fast, Accurate DNN Inference on Low-Cost Edges Using Heterogeneous Accelerator Execution},
year = {2025},
isbn = {9798400711961},
publisher = {Association for Computing Machinery},
address = {New York, NY, USA},
url = {https://doi.org/10.1145/3689031.3696067},
doi = {10.1145/3689031.3696067},
booktitle = {Proceedings of the Twentieth European Conference on Computer Systems},
pages = {507–523},
numpages = {17},
location = {Rotterdam, Netherlands},
series = {EuroSys '25}
}

@inproceedings{NN-Stretch,
author = {Wei, Jianyu and Cao, Ting and Cao, Shijie and Jiang, Shiqi and Fu, Shaowei and Yang, Mao and Zhang, Yanyong and Liu, Yunxin},
title = {NN-Stretch: Automatic Neural Network Branching for Parallel Inference on Heterogeneous Multi-Processors},
year = {2023},
isbn = {9798400701108},
publisher = {Association for Computing Machinery},
address = {New York, NY, USA},
url = {https://doi.org/10.1145/3581791.3596870},
doi = {10.1145/3581791.3596870},
booktitle = {Proceedings of the 21st Annual International Conference on Mobile Systems, Applications and Services},
pages = {70–83},
numpages = {14},
location = {Helsinki, Finland},
series = {MobiSys '23}
}

@inproceedings{10.1145/3372224.3419192,
author = {Yi, Juheon and Lee, Youngki},
title = {Heimdall: mobile GPU coordination platform for augmented reality applications},
year = {2020},
isbn = {9781450370851},
publisher = {Association for Computing Machinery},
address = {New York, NY, USA},
url = {https://doi.org/10.1145/3372224.3419192},
doi = {10.1145/3372224.3419192},
booktitle = {Proceedings of the 26th Annual International Conference on Mobile Computing and Networking},
articleno = {35},
numpages = {14},
location = {London, United Kingdom},
series = {MobiCom '20}
}

@inproceedings{10.1145/3498361.3538948,
author = {Jeong, Joo Seong and Lee, Jingyu and Kim, Donghyun and Jeon, Changmin and Jeong, Changjin and Lee, Youngki and Chun, Byung-Gon},
title = {Band: coordinated multi-DNN inference on heterogeneous mobile processors},
year = {2022},
isbn = {9781450391856},
publisher = {Association for Computing Machinery},
address = {New York, NY, USA},
url = {https://doi.org/10.1145/3498361.3538948},
doi = {10.1145/3498361.3538948},
booktitle = {Proceedings of the 20th Annual International Conference on Mobile Systems, Applications and Services},
pages = {235–247},
numpages = {13},
location = {Portland, Oregon},
series = {MobiSys '22}
}

@INPROCEEDINGS{9796661,
  author={Fu, Ziyan and Ren, Ju and Zhang, Deyu and Zhou, Yuezhi and Zhang, Yaoxue},
  booktitle={IEEE INFOCOM 2022 - IEEE Conference on Computer Communications}, 
  title={Kalmia: A Heterogeneous QoS-aware Scheduling Framework for DNN Tasks on Edge Servers}, 
  year={2022},
  volume={},
  number={},
  pages={780-789},
  doi={10.1109/INFOCOM48880.2022.9796661}}

@misc{zou2025surveyrealtimeschedulingacceleratorbased,
      title={A Survey of Real-time Scheduling on Accelerator-based Heterogeneous Architecture for Time Critical Applications}, 
      author={An Zou and Yuankai Xu and Yinchen Ni and Jintao Chen and Yehan Ma and Jing Li and Christopher Gill and Xuan Zhang and Yier Jin},
      year={2025},
      eprint={2505.11970},
      archivePrefix={arXiv},
      primaryClass={cs.DC},
      url={https://arxiv.org/abs/2505.11970}, 
}

@inproceedings{10.1145/3676641.3716278,
author = {Tan, Xin and Jiang, Yimin and Yang, Yitao and Xu, Hong},
title = {Towards End-to-End Optimization of LLM-based Applications with Ayo},
year = {2025},
isbn = {9798400710797},
publisher = {Association for Computing Machinery},
address = {New York, NY, USA},
url = {https://doi.org/10.1145/3676641.3716278},
doi = {10.1145/3676641.3716278},
booktitle = {Proceedings of the 30th ACM International Conference on Architectural Support for Programming Languages and Operating Systems, Volume 2},
pages = {1302–1316},
numpages = {15},
location = {Rotterdam, Netherlands},
series = {ASPLOS '25}
}

@misc{lee2024exploreselectderiverecall,
      title={Explore, Select, Derive, and Recall: Augmenting LLM with Human-like Memory for Mobile Task Automation}, 
      author={Sunjae Lee and Junyoung Choi and Jungjae Lee and Munim Hasan Wasi and Hojun Choi and Steven Y. Ko and Sangeun Oh and Insik Shin},
      year={2024},
      eprint={2312.03003},
      archivePrefix={arXiv},
      primaryClass={cs.HC},
      url={https://arxiv.org/abs/2312.03003}, 
}

@misc{subramanian2025smalllanguagemodelsslms,
      title={Small Language Models (SLMs) Can Still Pack a Punch: A survey}, 
      author={Shreyas Subramanian and Vikram Elango and Mecit Gungor},
      year={2025},
      eprint={2501.05465},
      archivePrefix={arXiv},
      primaryClass={cs.CL},
      url={https://arxiv.org/abs/2501.05465}, 
}

@inproceedings{10.1145/3662006.3662059,
author = {Li, Xiang and Lu, Zhenyan and Cai, Dongqi and Ma, Xiao and Xu, Mengwei},
title = {Large Language Models on Mobile Devices: Measurements, Analysis, and Insights},
year = {2024},
isbn = {9798400706639},
publisher = {Association for Computing Machinery},
address = {New York, NY, USA},
url = {https://doi.org/10.1145/3662006.3662059},
doi = {10.1145/3662006.3662059},
booktitle = {Proceedings of the Workshop on Edge and Mobile Foundation Models},
pages = {1–6},
numpages = {6},
location = {Minato-ku, Tokyo, Japan},
series = {EdgeFM '24}
}

@Misc{snpe,
	title           = {Qualcomm Neural Processing Engine.},
	howpublished    = {\url{https://docs.qualcomm.com/bundle/publicresource/topics/80-70015-15BY/snpe.html}},
        year = 2025
}

\end{document}